\documentclass[aps,prb,twocolumn,showpacs,amsmath]{revtex4}
\usepackage{graphicx}
\usepackage{bm}

\begin{document}

\title{Theory of flux cutting and flux transport at the critical current of a type-II superconducting cylindrical wire}

\author{John R.\ Clem}
\affiliation{%
   Ames Laboratory and Department of Physics and Astronomy,\\
  Iowa State University, Ames, Iowa 50011--3160, USA}

\date{\today}

\begin{abstract}
I introduce a critical-state  theory incorporating both flux cutting and flux transport to calculate the magnetic-field and current-density distributions inside a type-II superconducting cylinder at its critical current in a longitudinal applied magnetic field.  The theory is an extension of the elliptic critical-state model introduced by Romero-Salazar and P\'erez-Rodr\'iguez. The vortex dynamics depend in detail upon two nonlinear effective resistivities for flux cutting ($\rho_\parallel$) and flux flow ($\rho_\perp$), and their ratio $r= \rho_\parallel/\rho_\perp$. When $r < 1$, the low relative efficiency of flux cutting  in reducing the magnitude of the internal magnetic-flux density leads to a paramagnetic longitudinal magnetic moment.  As a model for understanding the experimentally observed interrelationship between the critical currents for flux cutting and depinning, I calculate the forces on a helical vortex arc stretched between two pinning centers when the vortex is subjected to a current density of arbitrary angle $\phi$.   Simultaneous initiation of flux cutting and flux transport occurs at the critical current density  $J_c(\phi)$ that makes the  vortex arc unstable.
\end{abstract}


\maketitle

\section{Introduction\label{Intro}}

The behavior of type-II superconductors carrying current in a perpendicular applied magnetic field is well understood in terms of the critical-state theory, first introduced by Bean.\cite{Bean62,Bean64} The fundamental idea is that the penetration of magnetic fields in the form of quantized vortices\cite{Abrikosov57} into practical type-II superconductors is impeded by an array of pinning centers.\cite{Campbell72}   The magnetic-flux distribution can be put into a variety of metastable states, and changes in the flux distribution can occur only when the magnitude of the local current density $\bm J = \nabla \times \bm H$ exceeds the critical-current density $J_c$, the threshold for depinning and {\it flux transport}.  The driving force per unit volume on a vortex array carrying magnetic flux density $\bm B$ is $\bm F = \bm J \times \bm B$, and the vortices move whenever the magnitude of this force locally exceeds the average pinning force per unit volume, $F_p = J_c B$.  Motion of the vortices with a local velocity $\bm v$ gives rise to a local flux-transport electric field\cite{Josephson65} $\bm E = \bm B \times \bm v$ perpendicular to $\bm B$.  This critical-state theory has served us well in providing a basis for practical applications of superconductivity, such as in magnet technology,\cite{Wilson83} where the magnetic fields generated are practically all perpendicular to the currents that generate the fields.  In addition, this theory has permitted a good understanding of the ac losses in many electric power applications,\cite{Carr83} since the local rate of power dissipation $\bm J \cdot \bm E$ is easily calculable when $J = |\bm J|$ is just above $J_c$.  

On the other hand, the corresponding theory for the behavior of type-II superconductors carrying current in a parallel applied magnetic field or in a field at an arbitrary angle relative to the current is not well developed.  With respect to the standard critical-state theory, two key questions are: (a) If $\bm J$ is not perpendicular to $\bm B$, how large can the (force-free) component of $\bm J$ parallel to $\bm B$ be?  Is there any limit to this component below the depairing current density?  (b) Superconducting wires subjected to a parallel magnetic field experimentally exhibit a finite electric field $\bm E$ with a component parallel to $\bm B$ when the current carried by the wire exceeds the critical current.\cite{Walmsley72a}  How is this component of the electric field produced?  {\it Flux cutting}\cite{Campbell72,Walmsley72a} provides natural answers to both of these questions: When $\bm J_\parallel$ (the component of $\bm J$ parallel to $\bm B$) is small, $\bm E_\parallel$ (the component of $\bm J$ parallel to $\bm B$) is zero, but when the magnitude of $\bm J_\parallel$ exceeds the threshold for flux cutting, flux-cutting processes initiated by local helical instabilities\cite{Clem77,Brandt81a,Brandt81b,Genenko94a,Genenko94b,Genenko95a,Genenko95b,Genenko96}   generate a finite value of  $\bm E_\parallel$.

Analogous processes occur in rotating superfluid $^4$He, where thermal counterflow parallel to the vortices produces turbulence initiated by the Glaberson-Donnelly helical instability.\cite{Glaberson74}  The resulting energy input to the vortex system is  dissipated at the microscopic level by vortex-line reconnection (the analogue of flux-line cutting),  recently filmed by Paoletti {\it et al}, \cite{Paoletti10} who analyzed the trajectories of solid hydrogen tracers in the superfluid to  identify tens of thousands of individual reconnection events between quantized vortices. 

 An extension of critical-state theory is needed to provide the theoretical basis for calculating ac losses in superconducting power transmission cables fabricated from multiple helically wound layers of second-generation YBCO tapes.\cite{Clem10}  The helical currents generate longitudinal magnetic fields inside the cable, such that the supercurrent density $\bm J$ has components both perpendicular and parallel to $\bm B$.

In this paper, I extend critical-state theory to account for both flux transport  and flux cutting in type-II superconductors.    In Sec.\ \ref{Basic}, I set down several of the basic equations needed and define the parallel and perpendicular components of $\bm J$ and $\bm E$.   I use capital letters to denote macroscopic fields of practical interest, $\bm B$, $\bm H$, $\bm J$, and $\bm E$, which in general depend on position $\bm r$ and time $t$.  These fields are local averages over a length scale several times the characteristic mesoscopic lengths of type-II superconductivity, namely the penetration depth $\lambda$, the coherence length $\xi$, and the intervortex spacing $a \sim \sqrt{\phi_0/B}$.  It has been known for over 50 years that in type-II superconductors the spatial variation of the mesoscopic fields $\bm b$, $\bm h$, $\bm j$, and $\bm e$ over these mesoscopic length scales is determined by vortices, which carry magnetic flux quantized in units of  $\phi_0 = h/2e$, the superconducting flux quantum.\cite{Abrikosov57}  I will focus here on using an extended critical-state theory to calculate the magnetic-field and current-density distributions just above the critical current of a type-II superconducting cylinder, where $\bm B$, $\bm H$, $\bm J$, and $\bm E$ are time-independent.  However, since $\bm E$ is here produced by dynamic processes at the mesoscopic length scale involving flux transport and flux cutting, we should keep in mind that $\bm b$, $\bm h$, $\bm j$, and $\bm e$ are all time-dependent quantities.

In Sec.\ \ref{NoHa}, I review how to use the standard critical-state theory to examine $\bm B$, $\bm H$, $\bm J$, and $\bm E$ at the critical current $I_c$ of the cylinder in zero applied magnetic field.  In Sec.\ \ref{Ha}, I show how to use an extended critical-state theory to calculate the magnetic-field and current-density distributions at $I_c$ in an applied longitudinal magnetic field $H_a$.  In Sec.\ \ref{Bdependence}, I calculate the dependence of $I_c$ upon $H_a$ when the critical current densities at the thresholds for flux transport and flux cutting depend upon the local value of $B$.  In  Sec.\ \ref{HelicalArcs}, I show how an applied current density affects the stability of a helical vortex arc stretched between two strong pinning centers.  The results provide us with a new model for the angular dependence of the critical current density, which simultaneously initiates flux cutting and flux transport.  In Sec.\ \ref{Discussion}, I discuss how the theoretical results qualitatively explain a variety of experiments, and I then turn to the issues of force-free configurations, the interactions between flux cutting and flux depinning, and the extensions needed to treat time-dependent problems.  

\section{Basic equations\label{Basic}}

\begin{figure}
\includegraphics[width=6cm]{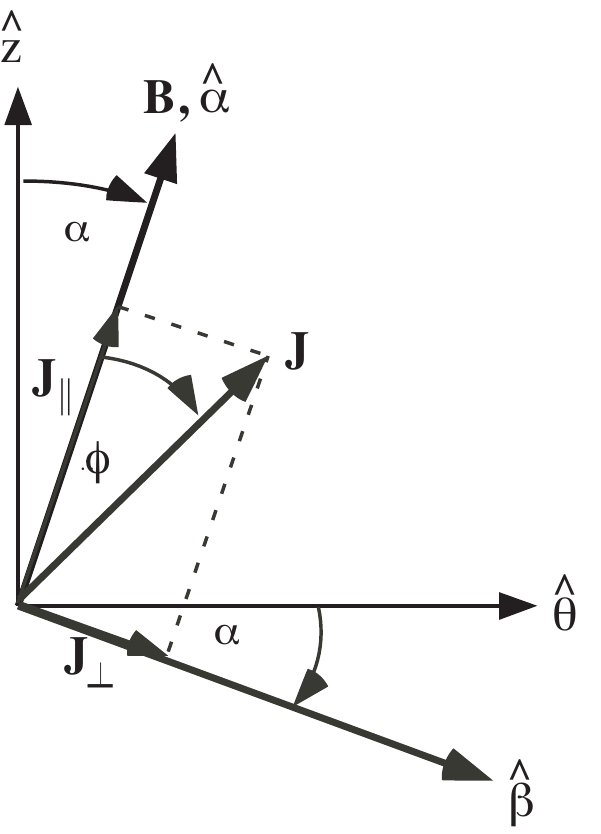}
\caption{%
Vectors used in this paper: The unit vectors in cylindrical coordinates are $\hat \rho$ (out of the paper), $\hat \theta$ (azimuthal), and $\hat z$ (longitudinal).  The local magnetic induction $\bm B = B \hat \alpha$ winds helically around the $z$ axis with pitch angle $\alpha$. The local current density $\bm J$ also winds helically around the $z$ axis but at an angle $\phi$ relative to $\bm B$.  The components of $\bm J$ along $\hat \alpha$ and $\hat \beta = \hat \alpha \times \hat \rho$ are $\bm J_\parallel$ and $\bm J_\perp$.  }
\label{Vectors}
\end{figure}

To describe the behavior at the critical current of a long type-II cylindrical wire of radius $R$ subjected to a parallel applied magnetic field $H_a$, let us assume that the local magnetic flux density $\bm B = B(\rho) \hat \alpha(\rho)$, where $\hat \alpha = \hat \theta \sin \alpha + \hat z \cos \alpha$, winds helically around the $z$ axis with a pitch angle $\alpha(\rho)$, as shown in Fig.\ \ref{Vectors}.  Assume also that the local current density $\bm J$ winds helically around the $z$ axis but at an angle $\phi(\rho)$ relative to $\bm B$.  The components of $\bm J$ along $\hat \alpha$ and $\hat \beta = \hat \alpha \times \hat \rho$ are $\bm J_\parallel = J_\parallel(\rho)\hat \alpha(\rho)$ and $\bm J_\perp= J_\perp(\rho)\hat\beta(\rho)$, where $\beta = \hat \theta \cos \alpha - \hat z \sin \alpha$. Similarly, when an electric field is generated, $\bm E = E_\parallel \hat \alpha + E_\perp \hat \beta.$

In the steady state, when there is no time dependence of $\bm B$ and $\bm J$, Ampere's law and Faraday's law require that 
\begin{eqnarray}
&&\bm J = \nabla \times \bm H, \label{Jeq}\\
&&\nabla \times \bm E = 0, \label{Eeq}
\end{eqnarray}
where $\bm H = (1/\mu_0)\nabla_{\bm B} F(\bm B)$ and $F(\bm B)$ is the Helmholtz free energy density in the superconducting state.\cite{Fetter69}

\section{Flux transport  at the critical current in zero applied longitudinal field\label{NoHa}}

We assume that the superconducting wire contains a randomly distributed dense array of point pinning centers.  We also consider the case that the wire initially is in the flux-free state, such that when a current $I$ is applied along the axis in the $z$ direction in the absence of an applied magnetic field, azimuthal magnetic flux penetrates in from the surface at $\rho = R$ with a distribution governed by a balance between the Lorentz force density $\bm J \times \bm B$ and the pinning force density $\bm F_p = \hat \rho J_{c\perp}B$.  In this case, wherever $\bm B \ne 0$, $\alpha = \pi/2$, $\hat \alpha = \hat \theta,$ $\hat \beta = -\hat z$, $\phi = 0$, $\bm J_\parallel = 0$, and $\bm J_\perp = -J_\perp \hat z = J_{c\perp} \hat z$. We usually expect $J_{c\perp}$ to be a monotonically decreasing function of $B$, but for the present let us use the Bean model, for which $J_{c\perp}$ is independent of $B$.

If we carefully distinguish between $\bm B$ and $\mu_0 \bm H$, we note that since $H(R) = I/2\pi R$, no vortices can enter the cylinder so long as $I < I_{c1},$ where $I_{c1} = 2\pi R H_{c1}$, where $H_{c1}$ is the lower critical field.  (We assume here that there are no surface barriers to vortex entry.)  If $J_{c\perp}< H_{c1}/R,$ the critical current is then $I_{c1}$, because any entering vortex ring of radius $\rho$ will simply collapse to zero radius under its own line tension, which is $\phi_0 H_{c1}/\rho$ per unit length.  

When $J_{c\perp}> H_{c1}/R,$ there is a minimum  radius, $\rho_c = H_{c1}/J_{c\perp} < R$, at which a vortex ring can be held by the pinning forces.  We seek an expression for the critical current $I_c$ at the first appearance of a steady-state longitudinal electric field, generated by the periodic  nucleation of vortex rings at the surface $\rho = R$ and their periodic self-annihilation when they are driven inwards to a radius slightly smaller than $\rho_c$. According to Faraday's law, Eq.\ (\ref{Eeq}), $\bm E$ is a constant, independent of $\rho$, which is given by the Josephson relation $\bm E = \bm B \times \bm v$.  Since here $\bm B = 
B_\theta \hat \theta$ and $\bm v = v_\rho \hat \rho$, the electric field has only a $z$ component, $E_z = -B_\theta v_\rho$.  (Note that $v_\rho = -|v_\rho|$, since vortices are constantly moving inward.) 

When $I_{c1} < I < I_c$, the solution of Eq.\ (\ref{Jeq}) is 
\begin{equation}
H(\rho) =\frac{I}{2\pi \rho}-\frac{(R^2-\rho^2)J_{c\perp}}{2\rho},\;\rho_{c1} \le \rho \le R,
\label{Htransport}
\end{equation}
where
\begin{equation}
\rho_{c1} = \rho_c + \sqrt{R^2+\rho_c^2-I/\pi J_{c\perp}}
\end{equation}
is the radius at which $H(\rho) = H_{c1}$ and $B(\rho) = 0$.  Note that $H_{c1} < H(\rho) < H(R)$ if $\rho_{c1} < \rho < R$.  The critical current $I_c$ is the value of $I$ at which $\rho_{c1}$ is reduced to $\rho_c$, such that
\begin{equation}
I_c = \pi (R^2+\rho_c^2)J_{c\perp},
\end{equation}
\begin{equation}
\frac{I_c}{I_{c1}} = \frac{R^2+\rho_c^2}{2R\rho_c},
\end{equation}
and 
\begin{equation}
H(\rho) =\frac{(\rho^2+\rho_c^2)J_{c\perp}}{2\rho},\;\rho_{c} \le \rho \le R.
\label{Htransportc}
\end{equation}

All the above results are greatly simplified if at the critical current  
$\rho_c = H_{c1}/J_{c\perp}\ll R$, which is equivalent to the condition that $H(R) = I_c/2\pi R \gg H_{c1}$.  In high-$\kappa$ superconductors it is well known that for $H \gg H_{c1}$, $B \approx \mu_0 H$ to good approximation.  Thus if one makes the approximation that $B = \mu_0 H$ from the outset and ignores line tension effects, this corresponds to setting $H_{c1}=0$ and $\rho_c = 0$ in the above equations.  When $0 < I < I_c$, $H(\rho)$ is then given by Eq.\ (\ref{Htransport}) but with $\rho_{c1} = \sqrt{R^2-I/\pi J_{c\perp}}$ denoting the radius of the penetrating flux front, where $B = 0$.  At the critical current, which becomes simply $I_c = \pi R^2 J_{c\perp}$, the field distribution is then $H(\rho) = \rho J_c/2$, and the current density is $J_z = J_{c\perp}$.  
For simplicity, we shall assume in the rest of this paper that at the critical current the magnitude of $\bm H$ at the surface is much larger than $H_{c1}$, such that the simplifying assumption $\bm B = \mu_0 \bm H$ is a good approximation.

\section{Flux transport and flux cutting at the critical current in an applied longitudinal field \label{Ha}}

\subsection{Extending the elliptic critical-state model}

While the behavior at the critical current in zero applied longitudinal magnetic field is described as above in terms of widely accepted critical state concepts, what happens at the critical current in a finite applied field is not yet well established theoretically.  I present here a theoretical description that I believe describes the fundamental physics of the behavior under these conditions.  

Let us assume that both an electrical current $I$ and a magnetic field $H_a$ are applied along the $z$ direction, parallel to the axis of  a long type-II cylindrical wire of radius $R$.  Under these conditions we can expect the magnetic induction $\bm B$ and the current density $\bm J$ to wind helically around the $z$ axis as indicated in Fig.\ \ref{Vectors}.  We need a number of equations to determine how the physical quantities vary with the radial coordinate $\rho$.  For simplicity, we make the simplifying approximation $\bm B =  \mu_0 \bm H$.  In cylindrical coordinates, Eq.\ (\ref{Jeq}) yields the two equations
\begin{eqnarray}
J_\perp&=&-\Big(\frac{dH}{d\rho} + \frac{H \sin^2\alpha}{\rho}\Big),\label{Jperp} \\
J_\parallel &=&H\Big(\frac{d\alpha}{d\rho} + \frac{\sin\alpha \cos\alpha}{\rho}\Big).\label{Jparallel}
\end{eqnarray}

For $H_a = 0$, we had $J_\parallel = 0$ and 
$\alpha = \pi/2$, so that Eq.\ (\ref{Jparallel}) was satisfied, and we had $J_\perp = -J_{c\perp}$, such that the solution of Eq.\ (\ref{Jperp})  for $H$ was given by Eq.\ (\ref{Htransport}).  

However, what determines the values of $J_\parallel$ and $J_\perp$ when $H_a > 0$?  One model, which we proposed in Refs.\ \onlinecite{Clem82,Clem84,Perez85a,Perez85b,Clem86} and has been called the generalized double-critical-state model (GDCSM), assumed that both  $|J_\parallel| = J_{c\parallel}$ and $|J_\perp| = J_{c\perp}$  at the critical current.  Here $J_{c\perp}$ was identified as the magnitude of $J_\perp$ at the threshold of flux transport (depinning), and $J_{c\parallel}$ was identified as the magnitude of $J_\parallel$ at the threshold of flux cutting. 
Recent experiments,\cite{Clem11} however, have found that two of the predictions of the GDCSM, a cusplike angular dependence of the critical-current density $J_c$ and a sawtoothlike behavior of the direction of the electric field $\bm E$ just above the critical current, are not seen experimentally.  The smooth angular dependence of $J_c$ observed experimentally\cite{Clem11,Herzog97,Durrell03,Rutter05,Durrell07,Maiorov06} is in much better agreement with an elliptic critical-state model, introduced by Romero-Salazar and P\'erez-Rodr\'iguez.\cite{Romero03a,Romero03b,Romero04}  However, the angular dependence of the electric field $\bm E$ for $J$ just above $J_c$ was found to require an extension of the elliptic critical-state model, to be described later.

\begin{figure}
\includegraphics[width=8cm]{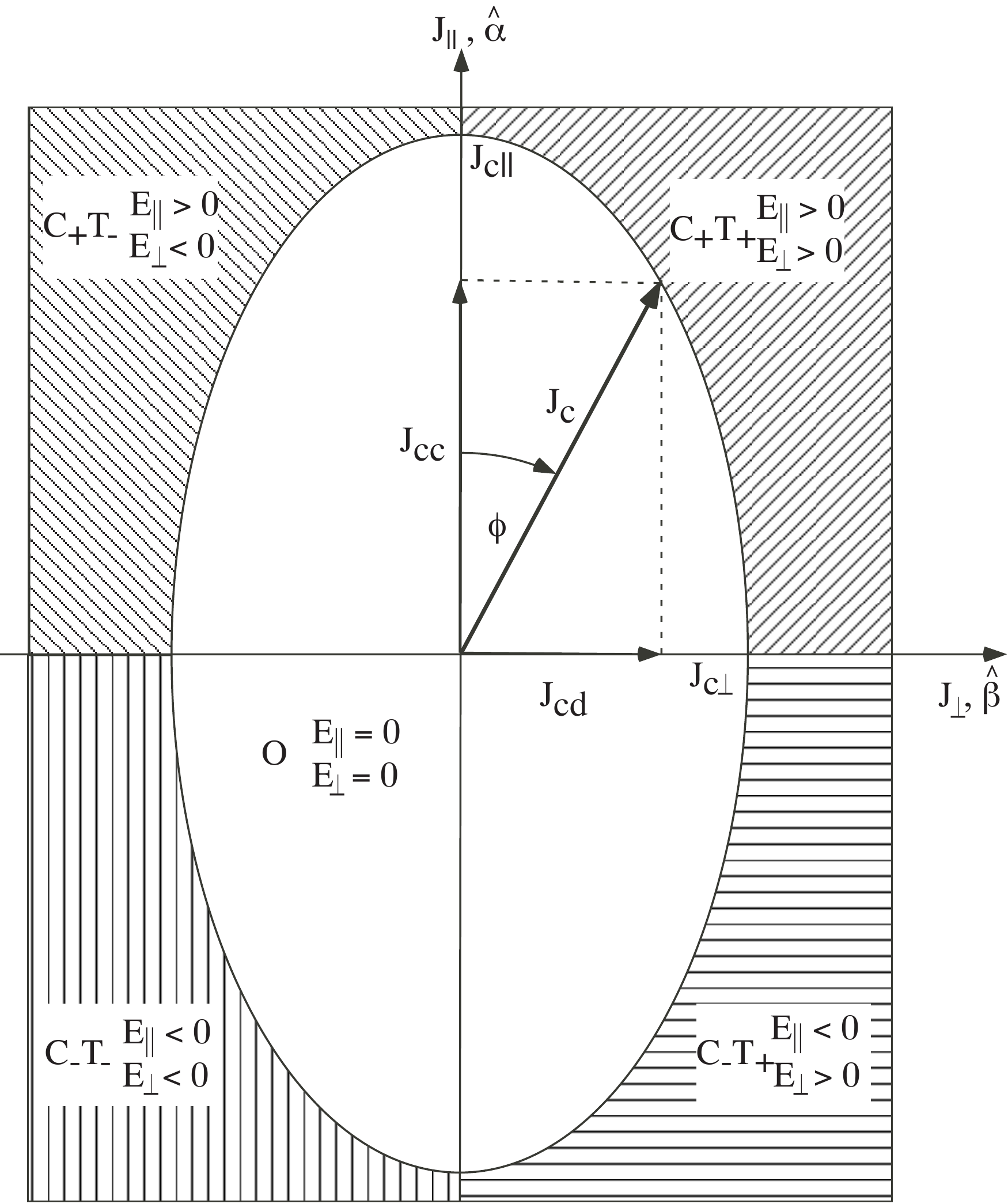}
\caption{Behavior of a vortex array as a function
of the magnitude $J$ and angle $\phi$ of the current density $\bm J$ relative to the direction of the flux density $\bm B = B \hat \alpha$ when
flux-line cutting and depinning interact and the critical current density $J_c(B,\phi)$ 
is given by the elliptic critical-state model of Refs.\
\onlinecite{Romero03a,Romero03b,Romero04}.}
\label{EllipticPlot}
\end{figure}

Figure \ref{EllipticPlot} shows the ellipse representing the critical current density $J_c(B,\phi)$ in the $J_\perp-J_\parallel$ plane, according to the original elliptic critical-state
model,\cite{Romero03a,Romero03b,Romero04}   where $\phi$ is the angle between $\bm J$ and $\bm B = B \hat \alpha$, and $J_{\perp} = J \sin \phi$ and $J_{\parallel} = J
\cos \phi$.
In the O zone inside the ellipse described by 
\begin{equation}
\frac{\sin^2\phi}{J_{c\perp}(B)^2}+
\frac{\cos^2\phi}{J_{c\parallel}(B)^2}=\frac{1}{J_c(B,\phi)^2}
\label{Jcellipse1}
\end{equation}
or 
\begin{equation}
J_c(B,\phi)=1/\sqrt{\frac{\sin^2\phi}{J_{c\perp}(B)^2}+
\frac{\cos^2\phi}{J_{c\parallel} (B)^2}},
\label{Jcellipse}
\end{equation}
neither flux transport (depinning) nor flux cutting occurs ($E_{\perp} = 0$ and
$E_{\parallel}=0$).
Flux transport, for which the vortices are
depinned and $|E_{\perp}| >  0$, occurs everywhere outside the ellipse (except when
$J_{\perp}=0$) in zones with labels including the symbol T$_+$ ($E_{\perp}>0$) or T$_-$ ($E_{\perp}<0$).  Flux-line cutting, for which
$|E_{\parallel}| >  0$, occurs everywhere outside the ellipse (except when
$J_{\parallel}=0$) in zones with labels including the symbol C$_+$  ($E_{\parallel} > 0$) or 
C$_-$ ($E_{\parallel} < 0$).  In other words, except for the special cases of $\phi = \pm \pi/2$, 0, and $\pi$, the critical current density for flux depinning and flux cutting is the same, $J_c(B,\phi)$.
When $\phi = \pm \pi/2$, flux transport (but no flux cutting) occurs when $J > J_c$, and when $\phi = 0$ or $\pi$, flux cutting (but no flux transport) occurs when $J > J_c$.

The magnitude of the component of the critical current density along $J_\perp$ associated with depinning (see Fig.\ \ref{EllipticPlot}) is 
\begin{equation}
J_{cd}(B,\phi) = J_c(B,\phi)|\sin\phi|.
\end{equation}
For fixed $J_\parallel$, the threshold for depinning is reached when $|J_\perp|$ increases to $J_{cd}$.  The magnitude of the component of the critical current density along $J_\parallel$ associated with cutting is \begin{equation}
J_{cc}(B,\phi) = J_c(B,\phi)|\cos\phi|.
\end{equation}
For fixed $J_\perp$, the threshold for  cutting is reached when $|J_\parallel|$ increases to $J_{cc}$.
In contrast to the assumptions of the GDCSM, important new features of the underlying physics within the elliptic critical-state
model are the  assumptions that  (a) the threshold
$J_{cd}$ for depinning is reduced as the magnitude of 
$J_{\parallel}$  increases, i.e., we can think of $J_{cd}$ is a monotonically decreasing function of the magnitude of 
$J_{\parallel}$, and (b) the threshold
$J_{cc}$ for flux-line cutting is reduced as 
$J_{\perp}$  increases, i.e., we can think of $J_{cc}$ is a monotonically decreasing function of the magnitude of
$J_{\perp}$.     
 
I propose that the above relationships between the thresholds for depinning and flux cutting are an essential general property of type-II superconductors, even if the mathematical form of the experimentally determined $J_c(\phi)$ deviates from the ellipse of Eq.\ (\ref{Jcellipse}) in a particular material.  In other words, (a) the threshold
$J_{cd}$ for depinning is always reduced to zero as the magnitude of 
$J_{\parallel}$ increases to its maximum threshold value
$J_{c\parallel}$ and (b) the threshold
$J_{cc}$ for flux-line cutting is always reduced to zero as 
$J_{\perp}$ increases to its maximum threshold value 
$J_{c\perp}$.  The underlying reason for this interrelationship is that when $\bm J$ is at an angle relative to $\bm B$, the helical instability is the triggering mechanism that results in both depinning and flux cutting. A simple model calculation for the critical current for the helical instability of an isolated vortex subsequently leading to depinning and flux cutting is given in Sec.\   \ref{HelicalArcs}.

In summary, at the critical current density of the elliptic critical-state
model, $J = J_c(B,\phi)$, and $\bm J$
lies on the ellipse of  Eq.\ (\ref{Jcellipse}), such that $J_{\perp} = J_c(B,\phi)
\sin
\phi$,  $J_{\parallel} = J_c(B,\phi)
\cos \phi$, and 
\begin{equation} 
J_c(B,\phi) = \frac{J_{c\perp}(B)}{\sqrt{\sin^2 \phi+\tan^2
\phi_c \cos^2
\phi}},
\label{Jcelliptic}
\end{equation} 
where $\tan \phi_c = J_{c\perp}(B)/J_{c\parallel}(B)$.
Note that within this model $J_{c\perp}(B) = J_c(B,\pi/2)$ and $J_{c\parallel}(B) = J_c(B,0)$.

Additional relations are needed to connect the components of $\bm J$ and $\bm E$.  Here we extend the original elliptic critical-state theory\cite{Romero03a,Romero03b,Romero04} by introducing the general relations
\begin{eqnarray}
E_\perp & = & \rho_\perp J_\perp, \label{Eperpelliptic}\\
E_\parallel & = &\rho_\parallel J_\parallel, \label{Eparallelelliptic} 
\end{eqnarray}
where  $\rho_\perp$ and $\rho_\parallel$  are nonlinear effective resistivities, which are expected to depend upon the current density $J$ and magnetic flux density $B$. Since the physics of flux cutting is different from that of flux depinning, we expect that $\rho_\parallel$ is not the same as $\rho_\perp$, and hence we do not expect $\bm E$ to be parallel to $\bm J$ in  general.  As shown below, a useful parameter is the ratio $r = \rho_\parallel/\rho_\perp$, which has been found experimentally to be independent of $J$ just above $J_c$.\cite{Clem11}

One could use the following model for the effective resistivities   $\rho_\perp$ and $\rho_\parallel$:  
\begin{eqnarray}
E_\perp &=&0, \; \;\;\;\;\;\;\;\;\;\;\;\;\;\;\;\;\;\;\;\;0 \le J_\perp\le J_{cd},\\
&=&\rho_d(J_\perp-J_{cd}),\;\;\;\;\;J_\perp> J_{cd},\\
E_\parallel &=&0, \; \;\;\;\;\;\;\;\;\;\;\;\;\;\;\;\;\;\;\;\;0 \le J_\parallel\le J_{cc}, \label{Eparallelzero}\\
&=&\rho_c(J_\parallel-J_{cc}),\;\;\;\;\;J_\parallel> J_{cc},
\end{eqnarray} 
such that $E = \sqrt{E_\perp^2+E_\parallel^2}$ and $J = \sqrt{J_\perp^2+J_\parallel^2}$ obey  
\begin{eqnarray}
E &=&0, \; \;\;\;\;\;\;\;\;\;\;\;\;\;\;\;\;\;\;\;\;\;\;\;\;J\le J_c(B,\phi),\\
&=&\rho[J-J_c(B,\phi)],\;\;\;J> J_c(B,\phi),\label{EvsJ}
\end{eqnarray}
where $\rho = \sqrt{\rho_d^2\sin^2\phi+\rho_c^2\cos^2\phi}$.  Here, $\rho_d$ corresponds to the flux-flow resistivity of Kim, Hempstead, and Strnad\cite{Kim64} and of Bardeen and Stephen,\cite{Bardeen65} and $\rho_c$ is a postulated analogous quantity.  
(The subscripts d and c refer to depinning and cutting.) 
The model of Eqs.\ (\ref{Eparallelzero})-(\ref{EvsJ}) is capable of describing the electric field components when there is a clear-cut linear onset of an electric field as $J$ crosses the ellipse $J_c(B,\phi)$ shown in Fig.\ \ref{EllipticPlot}.  On the other hand, if there are flux-creep effects that make the resistive transition more rounded, causing the critical-current density to depend upon an electric-field criterion, the more general model of Eqs.\ (\ref{Eperpelliptic}) and (\ref{Eparallelelliptic})   should be used.  

We are now ready to write down all the equations needed to calculate the current-density and magnetic-field distributions at the critical current of a cylinder.  Consistent with the usual critical-state approach for describing the behavior in the absence of a longitudinal field, we assume that just at the critical current we have  both  $|J_\perp| = J_{cd}$ and  $|J_\parallel| = J_{cc}$.  For a current $I$ and a magnetic field $H_a$ both applied in the $z$ direction we therefore have $J_\perp = J_c \sin \phi$, where $\phi < 0$, and  $J_\parallel = J_c \cos \phi$, which results in 
the two equations
\begin{eqnarray}
J_c \sin \phi&=&-\Big(\frac{dH}{d\rho} + \frac{H \sin^2\alpha}{\rho}\Big),\label{Jperpphi} \\
J_c \cos \phi&=&H\Big(\frac{d\alpha}{d\rho} + \frac{\sin\alpha \cos\alpha}{\rho}\Big).\label{Jparallelphi}
\end{eqnarray}

An additional equation is needed to relate $\alpha$ and $\phi$.  This comes from Eq.\ (\ref{Eeq}), which in cylindrical coordinates tells us that $\bm E = E_0 \hat z$, where $E_0 > 0$, so that $E_\perp = -E_0 \sin\alpha$ and $E_\parallel = E_0 \cos \alpha$.    Combining these equations with Eqs.\ (\ref{Eperpelliptic}) and (\ref{Eparallelelliptic}) yields
\begin{equation}
\tan \phi = -r \tan \alpha,
\label{tanphi}
\end{equation}
where $r = \rho_\parallel/\rho_\perp$.
We see immediately for the special case of $r = 1$ that we would have $\phi = -\alpha$, such that (see Fig.\ \ref{Vectors}) $\bm J$ would have only a $z$ component; Eq.\  (\ref{Jeq}) then tells us that the solution for $\bm H$ would have the property that $H_z(\rho) = H_a$.  For arbitrary values of $r$, however, we can use Eq.\   (\ref{tanphi}) to eliminate $\phi$ in favor of $\alpha$, which yields the following first-order differential equations for $H$ and $\alpha$:
\begin{eqnarray}
\frac{dH}{d\rho}\!\!&=& \!\!- \frac{H \sin^2\alpha}{\rho}
\!+\!\frac{r\sin\alpha}{\sqrt{r^2\sin^2\alpha/J_{c\perp}^2\!+\!\cos^2\alpha/J_{c\parallel}^2}}, \label{Hprime}\\
\frac{d\alpha}{d\rho}\!\! &=&\!\!- \frac{\sin\alpha \cos\alpha}{\rho}
\!+\!\frac{\cos\alpha}{H\sqrt{r^2\sin^2\alpha/J_{c\perp}^2\!+\!\cos^2\alpha/J_{c\parallel}^2}}.\;\;\;\;\;\;\;\label{alphaprime}
\end{eqnarray}

I shall refer to the collection of Eqs.\ (\ref{Jcellipse1}), (\ref{Jperpphi}), (\ref{Jparallelphi}), and (\ref{tanphi}) as the {\it extended elliptical critical-state model}, since the new equation, Eq.\ (\ref{tanphi}), goes beyond what was assumed in Refs.\ \onlinecite{Romero03a,Romero03b,Romero04}.  Note, however, that these equations are specialized for cylindrical geometry and would need generalization for other geometries.

\subsection{$B$ consumption by flux cutting\label{Bconsumption}}

As was noted in Refs.\ \onlinecite{Campbell72}, \onlinecite{Walmsley72a} and \onlinecite{Clem75}, the steady-state time-averaged voltage produced along a current-carrying type-II superconductor in a longitudinal magnetic field cannot be described as a flux-flow voltage generated by an inward collapsing array of helical vortices in the absence of  flux cutting.  The reason for this is that if no flux cutting is occurring, the electric field is given by $\bm E = \bm B \times \bm v$,\cite{Josephson65,Kim64,Kim69} where $\bm B$ is the locally averaged magnetic flux density generated by an array of vortices moving with a velocity $\bm v$.  According to Faraday's law, $\partial \bm B/\partial t = -\nabla \times \bm E $, a helical array of vortices  continuously nucleating at the surface and moving inward with a velocity $\bm v$ while carrying a longitudinal component of $\bm B$ would produce an azimuthal component of the electric field, thereby leading to an ever-increasing longitudinal magnetic flux density. 

Flux cutting is the means by which the time derivative of $\bm B$ can be reduced to zero.\cite{Clem82}  Multiplying Faraday's law by the unit vector $\hat \alpha$ and making use of Eq.\ (\ref{Jparallel}), we obtain the following equation describing the time rate of increase of $B$, the magnitude of $\bm B$, in cylindrical geometry:
\begin{equation}
\frac{\partial B}{\partial t} = -\frac{\partial E_\perp}{\partial \rho} -\frac{E_\perp \cos^2\alpha}{\rho} -\frac{J_\parallel E_\parallel}{H}.
\label{dBdt}
\end{equation}
The first two terms on the right-hand side of this equation, which could be expressed as $(\partial B/\partial t)_{transport},$ is simply the rate at which the local value of $B$ is increased by the transport of $\bm B$ toward the cylinder axis with a velocity $\bm v = \bm E \times \bm B/B^2$.  For the experimental conditions considered here, $(\partial B/\partial t)_{transport}$ is always positive.  The third term on the right-hand side, which could be expressed as $(\partial B/\partial t)_{cutting},$ is the rate at which the local value of $B$ is increased as a consequence of flux-line cutting.  However, since $J_\parallel E_\parallel$ is the rate of energy dissipation per unit volume, we see that  $(\partial B/\partial t)_{cutting}$ is always negative wherever flux cutting is occurring.  In the steady state, when $\partial B/\partial t = 0$, the local rate of increase of $B$ due to flux transport is exactly balanced by the local rate of decrease of $B$ due to flux cutting, as can be shown with the help of Eqs.\ (\ref{Jperpphi})-(\ref{tanphi}).

According to the Josephson relation, in the presence of flux cutting, a uniform steady-state longitudinal electric field must be given by $E_0 = h \nu'/2e = \phi_0 \nu',$ where $\nu'$ is the rate per unit length with which azimuthally directed flux quanta move inward and intersect a line parallel to the cylinder axis.  Cyclic flux-cutting processes initiated by local helical instabilities\cite{Clem77,Brandt81a,Brandt81b,Genenko94a,Genenko94b,Genenko95a,Genenko95b,Genenko96} allow helical vortices of one pitch to enter the cylinder and an equal number of vortices of longer pitch to exit the cylinder in such a way that on the average there is no net change in the number of longitudinal flux quanta in the cylinder.  For this reason the time-averaged azimuthal component of the electric field is zero.

We have seen in the above discussion that there is a competition between flux transport, which tends to increase $B$ inside the cylinder, and flux cutting, which tends to decrease $B$.  The efficiency of cutting relative to transport, as represented by the ratio $r = \rho_\parallel/\rho_\perp$ [see 
 Eqs.\ (\ref{Eperpelliptic}) and (\ref{Eparallelelliptic})] determines whether $\overline {H_z}$, the average of $H_z$ over the sample volume, is greater than, equal to, or less than the longitudinal applied field $H_a$.  If $r < 1$, corresponding to a low efficiency of flux cutting, we obtain $\overline {H_z}> H_a$ and a paramagnetic longitudinal magnetic moment.  If $r = 1$, corresponding to equal efficiencies of flux cutting and flux transport, we obtain $\overline {H_z} = H_a$.  If $r >1$, corresponding to a high efficiency of flux cutting, we  obtain $\overline {H_z} < H_a$ and a diamagnetic longitudinal magnetic moment. 

\subsection{Sample calculations\label{calcs}}

I now present a few results of sample calculations of the magnetic-field and current-density distributions at the critical current of a type-II superconducting cylinder.  Let $H_a$ denote the longitudinal applied magnetic field, and assume that there is no surface barrier at the superconductor's surface at $\rho = R$, such that $H(R) \cos\alpha(R) = H_a$.  (For simplicity, we consider here only the case $H_a > 0$.)  By Ampere's law, at the critical current $I_c$ where a finite steady-state electric field $\bm E = E_0 \hat z$ first appears, we also have $H(R)\sin\alpha(R) = I_c/2\pi R$.  The behavior depends in important ways upon the value of   $r = \rho_\parallel/\rho_\perp < 1$ [see Eqs.\ (\ref{Eperpelliptic}) and (\ref{Eparallelelliptic})]. 

Depending upon the values of $r$ (assumed here to be independent of $J$ and $H$) and $H_a$, the field along the axis $H(0) =H_0$ can be either finite or zero. When $H_0 > 0$, we must have $\alpha(\rho)=k_\alpha \rho$ as $\rho \to 0$, where $k_\alpha = J_{c\parallel}/2H_0$, as can been seen from Eq.\ (\ref{alphaprime}).  On the other hand, if $H(0) = 0$, we may have $\alpha(0) = \alpha_0,$ where $0 < \alpha_0 \le  \pi/2$, but then we must have $H(\rho) = k_H \rho$ as $\rho \to 0$, where 
\begin{equation}
k_H = \frac{r \sin \alpha_0}{(1+\sin^2\alpha_0)\sqrt{r^2\sin^2\alpha_0/J_{c\perp}^2+\cos^2\alpha_0/J_{c\parallel}}},
\end{equation}
as can be seen from Eq.\ (\ref{Hprime}).

\subsubsection{No magnetic moment for $r = 1$}

\begin{figure}
\includegraphics[width=8cm]{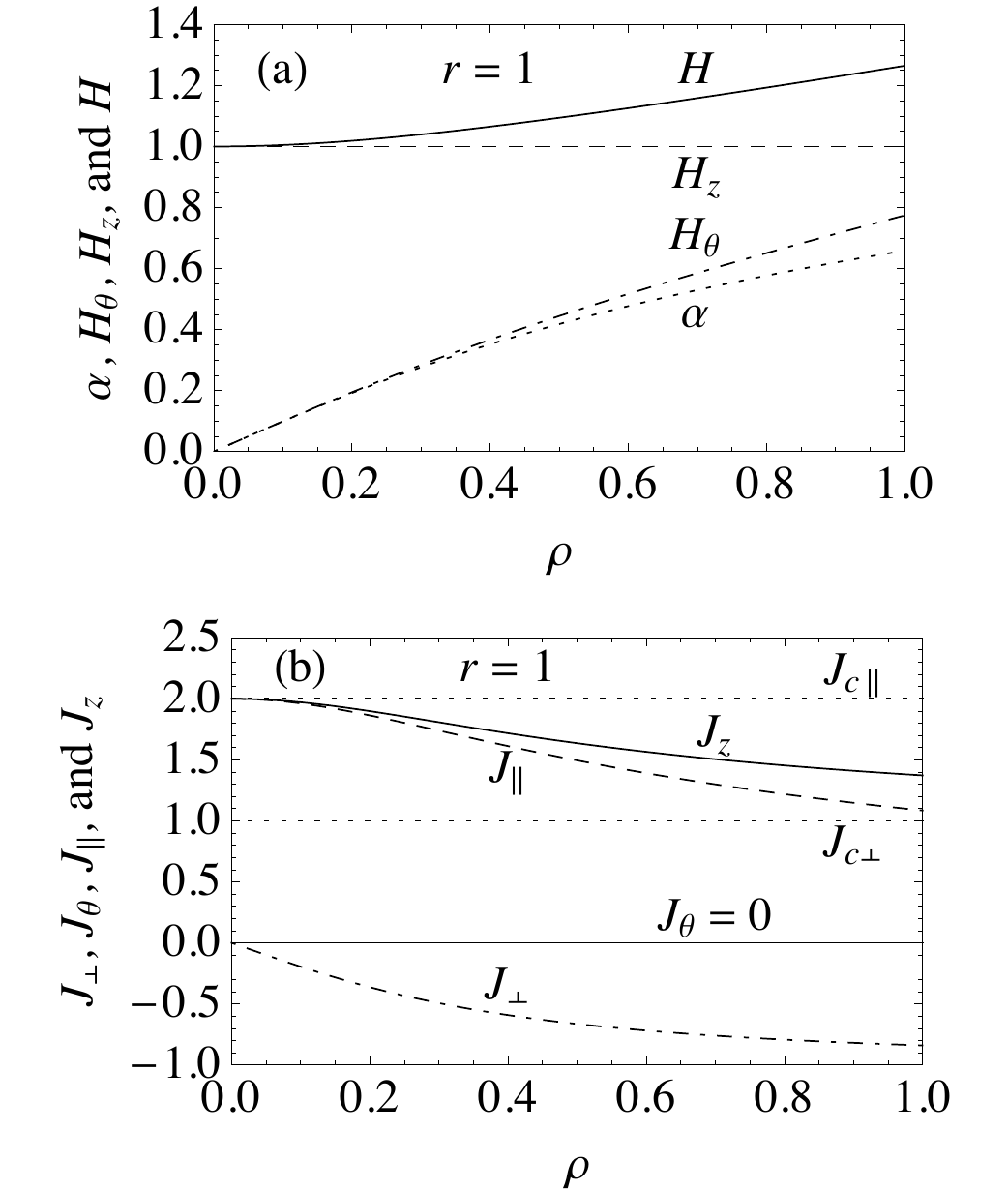}
\caption{Solutions of Eqs.\ (\ref{Hprime}) and (\ref{alphaprime}) vs $\rho$ for $r = 1$,  $H(0) = 1$, and $R = 1$. (a) $H$ (solid), $H_z$ (dashed), $H_\theta$ (dot-dashed), and $\alpha$ (dotted). (b) $J_z$ (solid), $J_\theta = 0$ (solid), $J_\parallel$ (dashed), and $J_\perp$ (dot-dashed) with the constants $J_{c\parallel} = 2$ and $J_{c\perp} = 1$ shown as dotted lines.}.
\label{HJPlotreq1}
\end{figure}

Numerical solutions of Eqs.\ (\ref{Hprime}) and (\ref{alphaprime}) for $r = 1$ are shown in Fig.\ \ref{HJPlotreq1}.  Note that $H_z$ is a constant, equal to the applied longitudinal field when $r = 1$, as discussed in Sec.\ \ref{Bconsumption}.  The reason for this is that when $r = 1$ [see Eqs.\ (\ref{Eperpelliptic}) and (\ref{Eparallelelliptic})], $\bm J$ is parallel to $\bm E$, and since $\bm E = E_0 \hat z$, $\bm J$ has only a $z$ component, such that $J_\theta = -dH_z(\rho)/d\rho = 0$ and $H_z = H_a$.  

\subsubsection{Paramagnetic moment for $r < 1$\label{para}}

\begin{figure}
\includegraphics[width=8cm]{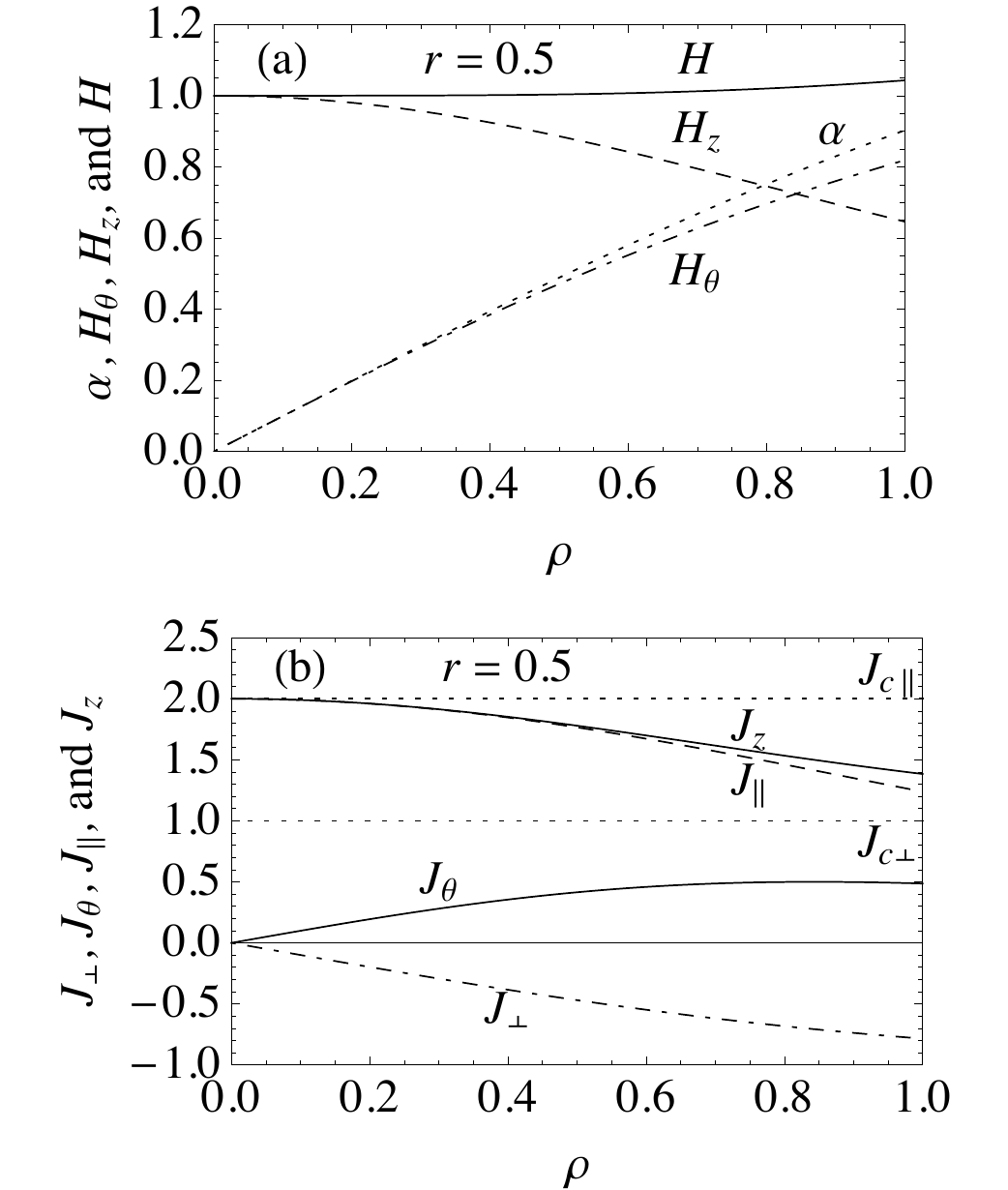}
\caption{Solutions of Eqs.\ (\ref{Hprime}) and (\ref{alphaprime}) vs $\rho$ for $r = 0.5$,  $H(0) = 1$, and $R = 1$. (a) $H$ (solid), $H_z$ (dashed), $H_\theta$ (dot-dashed), and $\alpha$ (dotted). (b) $J_z$ (solid), $J_\theta$ (solid), $J_\parallel$ (dashed), and $J_\perp$ (dot-dashed) with the constants $J_{c\parallel} = 2$ and $J_{c\perp} = 1$ shown as dotted lines.}.
\label{HJPlotreq0p5}
\end{figure}

Solutions for $r = 0.5$ are shown in Fig.\ \ref{HJPlotreq0p5}.  When $r < 1$, $\bm J$ is no longer parallel to $\bm E$, and as field lines of $\bm H$ wind around the $z$ axis as right-handed helices with a pitch angle $\alpha(\rho)>0$, streamlines of the current density $\bm J$ also wind around the $z$ axis as right-handed helices with a pitch angle [see Eq.\ \ref{tanphi}]
\begin{equation}
\psi_J=\alpha + \phi=\tan^{-1}\Big[\frac{(1-r)\tan\alpha}{1+r\tan^2\alpha}\Big].
\label{Jangle}
\end{equation}
Since $\tan \alpha >0$,  $\psi_J > 0$ when $r < 1$.   Since $J_\theta = -dH_z(\rho)/d\rho > 0$, $H_z$ decreases monotonically with increasing $\rho$, resulting in a paramagnetic magnetic moment per unit volume $M_z = \overline {H_z}-H_a > 0.$   As discussed in Sec.\ \ref{Bconsumption}, $B$ consumption due to flux cutting [see Eq.\ (\ref{dBdt})] is needed to prevent an ever-increasing buildup of longitudinal flux.  When $r = \rho_\parallel/\rho_\perp < 1$, flux cutting is less efficient in consuming $B$, and this allows a larger value of $\overline {H_z}$ in the steady state.

\begin{figure}
\includegraphics[width=8cm]{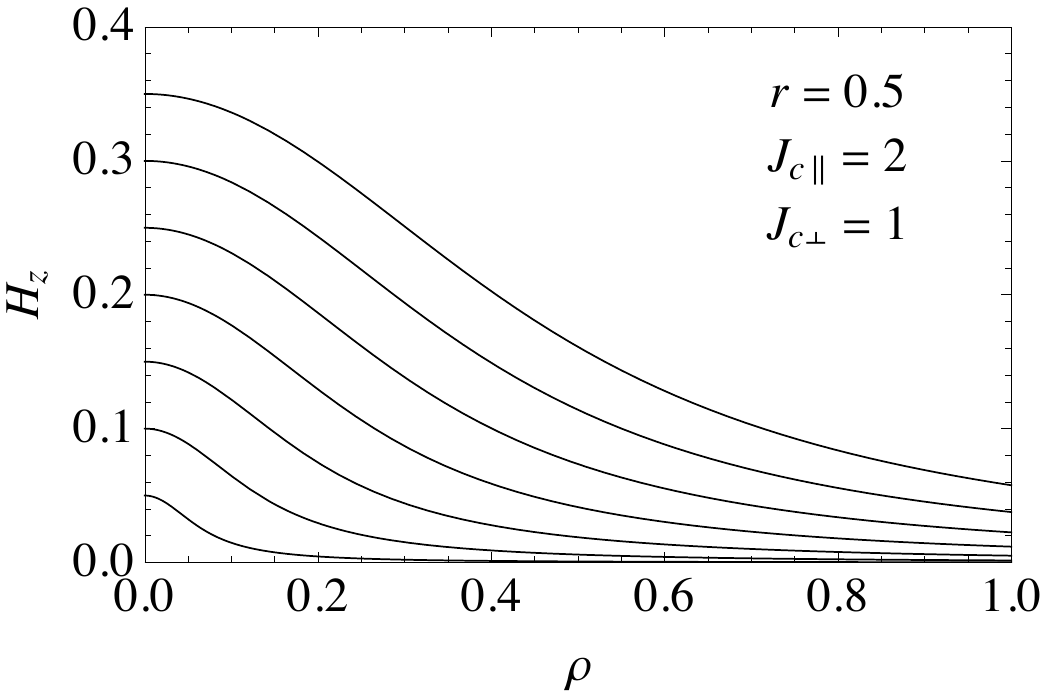}
\caption{$H_z$ calculated from Eqs.\ (\ref{Hprime}) and (\ref{alphaprime}) vs $\rho$ for $r = 0.5$, $J_{c\parallel} = 2$, $J_{c\perp} = 1$, and $R = 1$, showing a pronounced buildup of longitudinal flux along the cylinder axis even for very small values of the applied field $H_a = H_z(R)$.}.
\label{Hzvsrhoparamagplot}
\end{figure}

When $r < 1$, the application of even a very small longitudinal magnetic field $H_a$ can lead to a much larger value of the field $H_z$ along the axis of the cylinder.  Calculations illustrating this effect are shown in Fig.\   \ref{Hzvsrhoparamagplot}.

\subsubsection{Diamagnetic moment for $r > 1$}

\begin{figure}
\includegraphics[width=8cm]{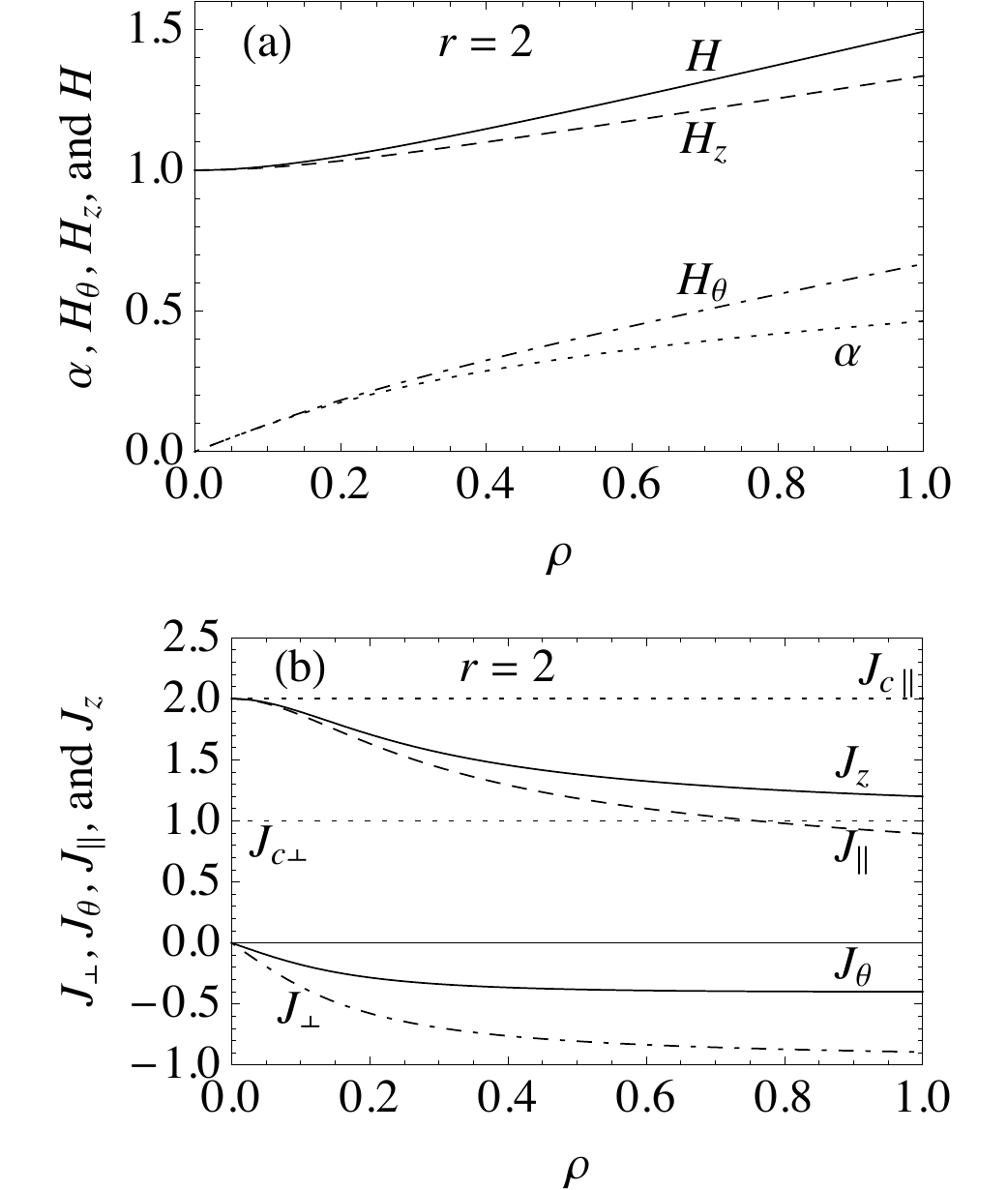}
\caption{Solutions of Eqs.\ (\ref{Hprime}) and (\ref{alphaprime}) vs $\rho$ for $r = 2$,  $H(0) = 1$, and $R = 1$. (a) $H$ (solid), $H_z$ (dashed), $H_\theta$ (dot-dashed), and $\alpha$ (dotted). (b) $J_z$ (solid), $J_\theta$ (solid), $J_\parallel$ (dashed), and $J_\perp$ (dot-dashed) with the constants $J_{c\parallel} = 2$ and $J_{c\perp} = 1$ shown as dotted lines.}.
\label{HJPlotreq2}
\end{figure}

Solutions for $r = 2$ are shown in Fig.\ \ref{HJPlotreq2}.  When $r > 1$, as field lines of $\bm H$ wind around the $z$ axis as right-handed helices with a pitch angle $\alpha(\rho)>0$, streamlines of the current density $\bm J$ wind around the $z$ axis as left-handed helices, i.e.,  with a negative pitch angle $\psi_J < 0$ [see Eq.\ (\ref{Jangle})].   Since $J_\theta = -dH_z(\rho)/d\rho < 0$, $H_z$ increases monotonically with increasing $\rho$, resulting in a diamagnetic magnetic moment per unit volume $M_z = \overline {H_z}-H_a < 0.$   As discussed in Sec.\ \ref{Bconsumption}, flux cutting [see Eq.\ (\ref{dBdt})] prevents an ever-increasing buildup of longitudinal flux.  When $r = \rho_\parallel/\rho_\perp > 1$, flux cutting is more efficient in consuming $B$ and this results in a smaller value of $\overline {H_z}$ in the steady state.

\begin{figure}
\includegraphics[width=8cm]{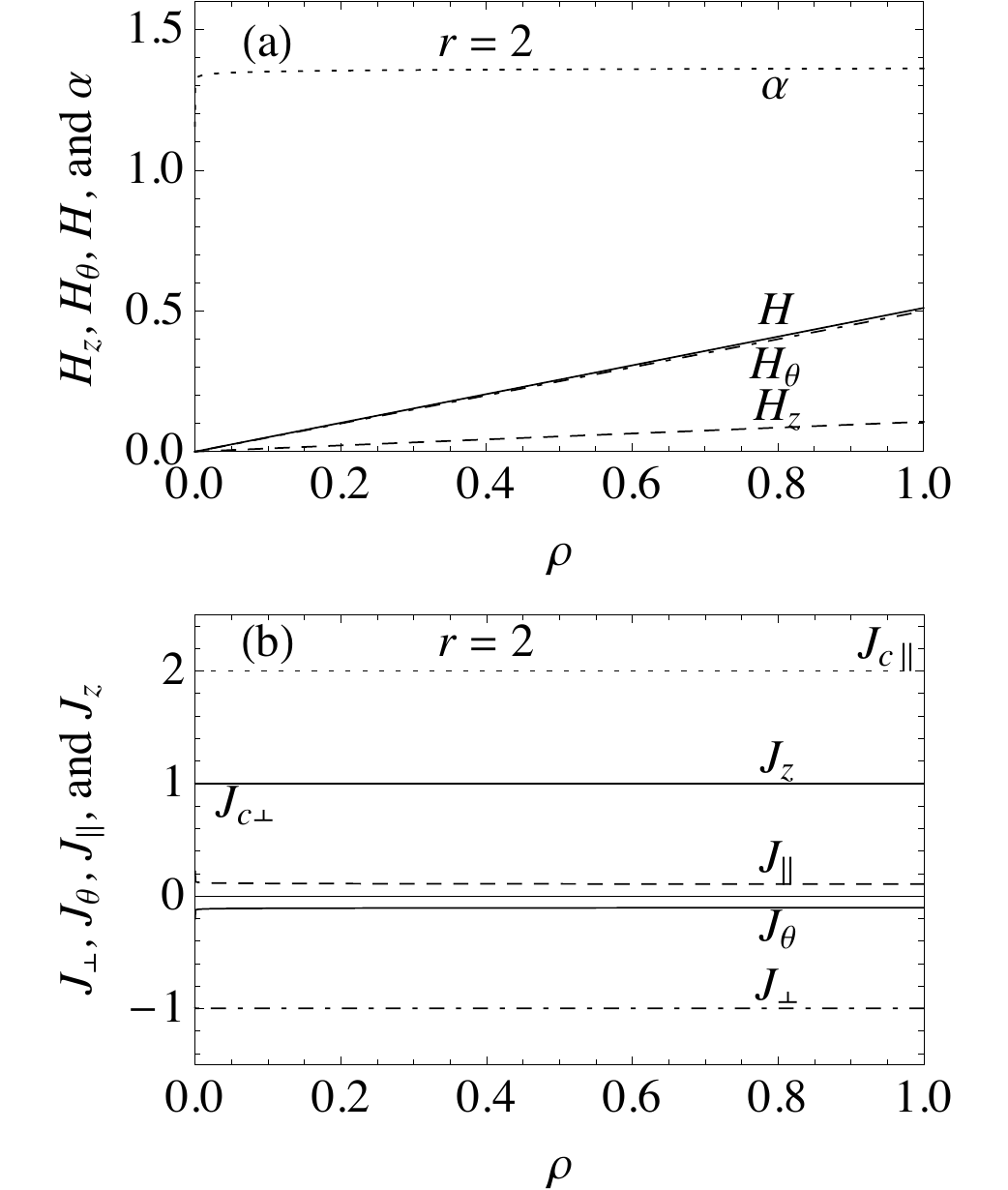}
\caption{Solutions of Eqs.\ (\ref{Hprime}) and (\ref{alphaprime}) vs $\rho$ for $r = 2$,  $H(0) = 0$, $\alpha_0 = \pi/4$, and $R = 1$. (a) $H$ (solid), $H_z$ (dashed), $H_\theta$ (dot-dashed), and $\alpha$ (dotted). (b) $J_z$ (solid), $J_\theta$ (solid), $J_\parallel$ (dashed), and $J_\perp$ (dot-dashed) with the constants $J_{c\parallel} = 2$ and $J_{c\perp} = 1$ shown as dotted lines.}.
\label{HJPlotreq2H0zero}
\end{figure}

\subsubsection{Critical current vs $H_a$}

\begin{figure}
\includegraphics[width=8cm]{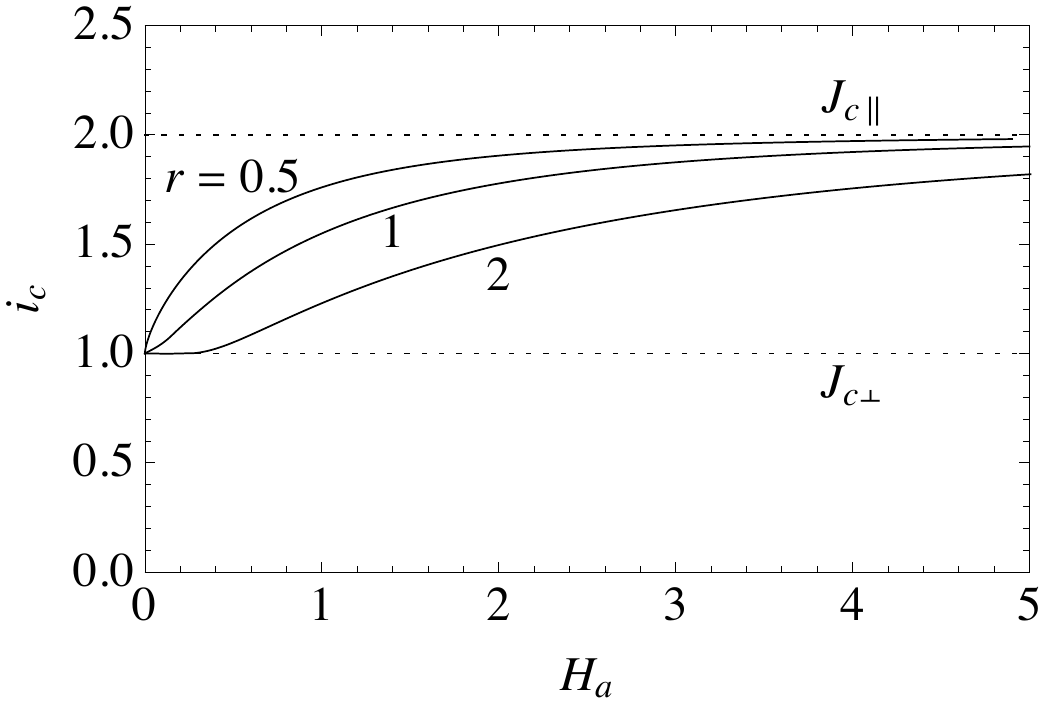}
\caption{ $i_c = I_c/I_{c\perp}$ (solid) vs $H_a$ for $r$ = 0.5, 1, and 2, calculated assuming $R = 1$, $J_{c\parallel} = 2$ (dotted), and $J_{c\perp} = 1$ (dotted), where $I_{c\perp} = \pi R^2 J_{c\perp}$ (see text).}.
\label{icvsHaplot}
\end{figure}

Shown in Fig.\ \ref{icvsHaplot} is the critical current $i_c = I_c/I_{c\perp}$, where $I_c = 2\pi R H_\theta(R)$ and $I_c = \pi R^2 J_{c\perp}$, as a function of $H_a = H_z(R)$, obtained Eqs.\ (\ref{Hprime}) and (\ref{alphaprime}), vs $\rho$ for several values of $r$ assuming $R = 1$, $J_{c\parallel} = 2$,  $J_{c\perp} = 1$, and no $B$ dependence of $J_c$ (Bean model). 

\begin{figure}
\includegraphics[width=8cm]{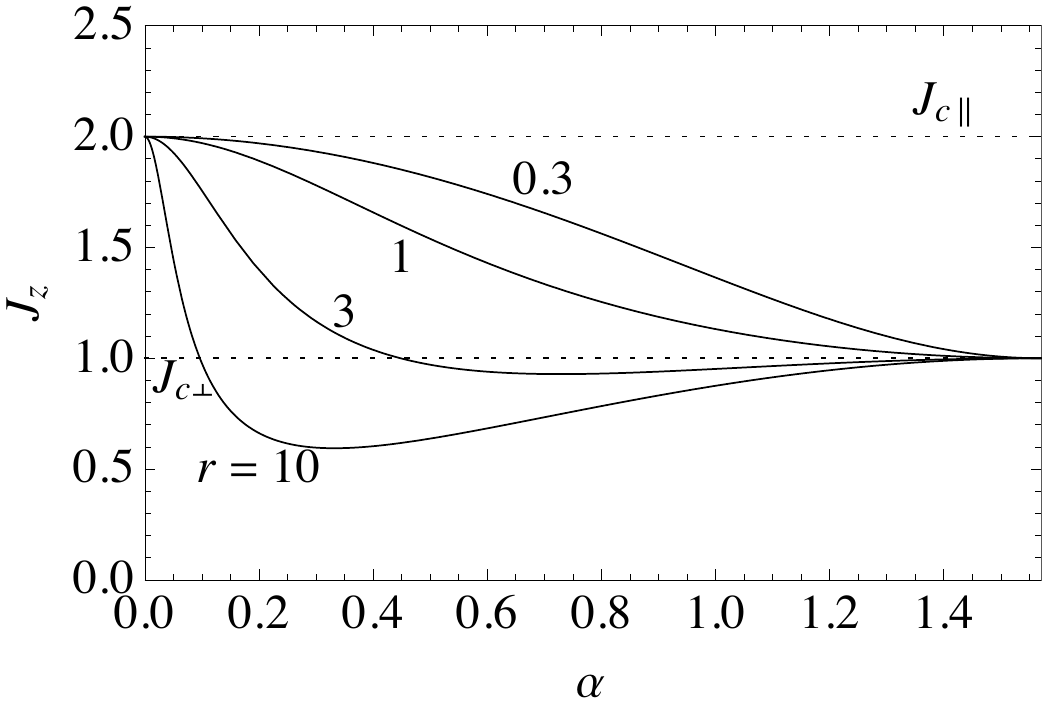}
\caption{$J_z$ [Eq.\ (\ref{Jzvsalpha})] vs $\rho$ for $r$ = 0.3, 1, 3, and 10 with the constants $J_{c\parallel} = 2$ and $J_{c\perp} = 1$ shown as dotted lines.}.
\label{Jzvsalphaplot}
\end{figure}

Surprisingly, numerical calculations reveal that, depending upon the value of $r$, the critical current $I_c = 2 \pi \int_0^R J_z(\rho) \rho d\rho$ can be less than $I_{c\perp} = \pi R^2 J_{c\perp}$.  (For example, when $r = 2$, $H_a$ = 0.113, $i_c$ = 0.9987, with all values of $i_c$ dipping slightly below 1 for small values of $H_a$ in Fig.\ \ref{icvsHaplot}.)  The reason for this is that $J_z = J_\parallel \cos \alpha - J_\perp \sin \alpha$, and at the critical current we have 
\begin{equation}
J_z = \frac{r \sin^2 \alpha + \cos^2 \alpha}{\sqrt{r^2\sin^2\alpha/J_{c\perp}^2\!+\!\cos^2\alpha/J_{c\parallel}^2}}.
\label{Jzvsalpha}
\end{equation}
As shown in Fig.\ \ref{Jzvsalphaplot}, $J_z$ can be less than $J_{c\perp}$ for some combinations of $r$ and $\alpha$, such that the integral over $\rho$ yielding $I_c$ can be less than $I_{c\perp}$, the effect being most pronounced for large values of $r$.

\section{Accounting for  field dependence of $J_{c\perp}$ and $J_{c\parallel}$\label{Bdependence}}

\begin{figure}
\includegraphics[width=8cm]{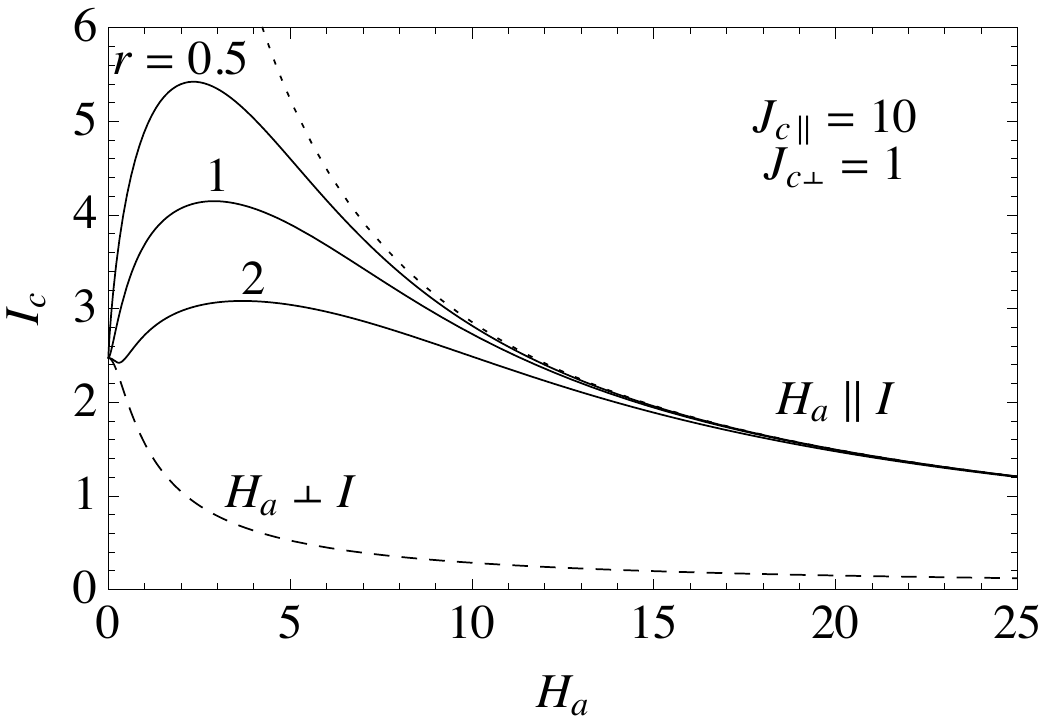}
\caption{ $I_c$ vs $H_a$ parallel to $I$  (solid) and $I_c$  vs $H_a$ perpendicular to $I$ (dashed) for $r$ = 0.5, 1, and 2, calculated assuming $R = 1$, $J_{c\parallel} = 10$, $J_{c\perp} = 1$, and Kim-model field dependence of $J_c$ with $H_0=1$ (see text). The dotted curve shows $\pi R^2 J_{c\parallel}/(1+H_a/H_0)$, while $\pi R^2 J_{c\perp}/(1+H_a/H_0)$ is indistinguishable from the dashed curve on this plot.}.
\label{IcvsHaKimplot}
\end{figure}

For simplicity, the above sample calculations were carried out using the assumption that  $J_c$ was a function of the angle $\phi$ between $\bm J$ and $\bm B$ but was a constant independent of $B$ or $H$, analogous to the commonly used Bean model.  However, because the solutions of Eqs.\ (\ref{Hprime}) and (\ref{alphaprime}) are obtained numerically, it is straightforward to incorporate field dependence into $J_c(B,\phi)$.  As an example, shown as the solid curves in Fig.\ \ref{IcvsHaKimplot} are sample plots of $I_c$ for $H_a$ applied parallel to the cylinder axis, calculated assuming that $J_c(B,\phi)$ has the same dependence as in the Kim model, $J_c(B) = J_c(0)/(1+B/B_0)$.  For simplicity, in this paper we have assumed $B = \mu_0 H$, so that for calculations of $I_c$ when a longitudinal field is applied, $J_{c\parallel}$ and $J_{c\perp}$ in Eqs.\ (\ref{Hprime}) and (\ref{alphaprime}) are replaced by their Kim-model analogs, $J_{c\parallel}/(1+H/H_0)$ and $J_{c\perp}/(1+H/H_0)$.  Note that self-field effects reduce  the critical current in zero applied field to the value $I_c(0) = 2.47$,  below the Bean-model result, $I_{c\perp} = \pi R^2 J_{c\perp}= 3.14$.  Although $I_c$ for parallel $H_a$ depends upon $r$ for relatively small fields, note that $I_c \approx \pi R^2 J_{c\parallel}/(1+H_a/H_0)$ for large $H_a$, independent of $r$, because in this case $\alpha(\rho)$ and $\phi(\rho)$ are both very close to zero for all $\rho$.

  When $H_a$ is applied perpendicular to the cylinder axis, $I_c(H_a)$ must reduce to its self-field value when $H_a = 0$ and to $I_c(H_a) \approx \pi R^2 J_{c\perp}/(1+H_a/H_0)$ when $H_a$ is much greater than the self-field. However, application of an arbitrary perpendicular field destroys the azimuthal symmetry of the field and current distributions, which now must depend on both the radial and azimuthal coordinates. Nevertheless, for simplicity, I have calculated the critical current using $\alpha = \pi/2$ and $H(\rho) = H_\theta(\rho)$ for $\rho < R$ obtained as the solution of  
\begin{equation}
\frac{dH}{d\rho}= - \frac{H}{\rho}
+\frac{J_{c\perp}}{1+\sqrt{H^2+H_a^2}/H_0}, \label{Hprimeapprox}
\end{equation}
an interpolation approximation that yields the correct $I_c$ in  both the self-field limit and the large-$H_a$ limit.

\section{Single-vortex model for $J_c$\label{HelicalArcs}}

\begin{figure}
\includegraphics[width=8cm]{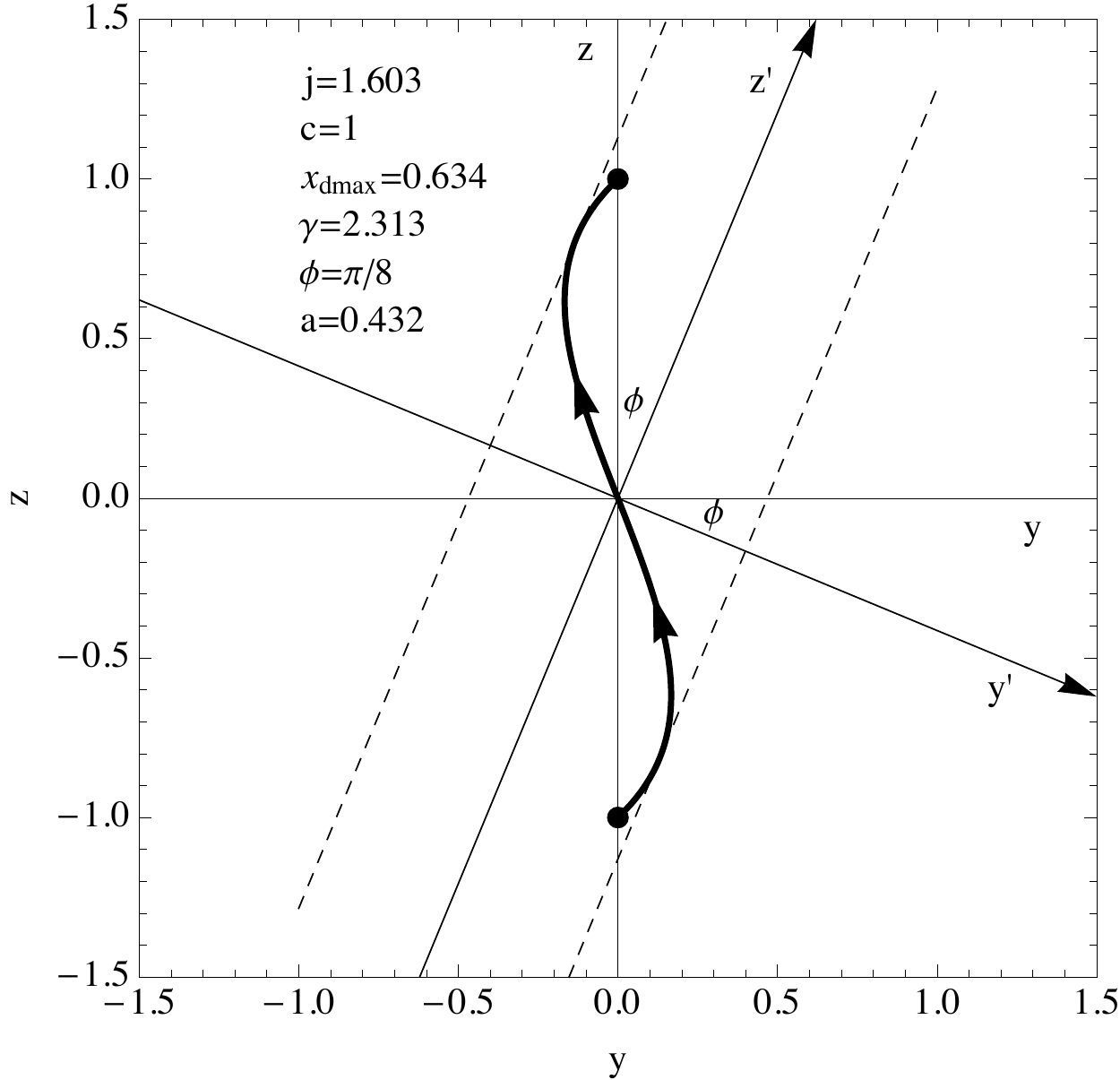}
\caption{ Helical vortex arc viewed looking down the $x$ axis, calculated at the critical current for the helical instability when $\phi=\pi/8$ and $c = 1$ (the corresponding parameters are shown in the inset).  The current density $\bm J$ is parallel to the $z'$ axis, and the helical arc wraps more than halfway around the cylinder of radius $a$ (dashed), which is centered on the $z'$ axis.  The $z'$ axis intersects the $x$ axis at $x=x_0 = x_{dmax}-a$ = 0.201.  The vortex arc intersects the $x$ axis at $x_{dmax} = 0.634$, and the ends of the arc are attached to strong pins at $\bm r_\pm = (0,0,\pm 1)$.}
\label{helixfig}
\end{figure}

I  next present a simple model that shows an intimate connection between the thresholds for depinning and flux cutting.  Consider for simplicity a vortex segment stretching between two very strong pinning centers at $\bm r_\pm =(x,y,z) = (0,0,\pm c)$  (see Fig.\ \ref{helixfig}). Assume that the vortex is subjected to a current density $\bm J = \hat y J_y + \hat z J_z$, where $J_y = J \sin \phi$ and $J_z = J \cos \phi$.  (For simplicity, we here consider behavior in a Cartesian coordinate system rather than the cylindrical coordinate system used in Fig.\ \ref{Vectors}.  However, in both cases we may think of the $z$ axis  as the vortex direction, and $\phi$ as the angle of $\bm J$ away from the vortex direction, as in Figs.\ \ref{Vectors} and \ref{EllipticPlot}.)   Since the Lorentz force per unit length of magnitude $J_y\phi_0$ due to the component $J_y$ is perpendicular to the vortex line, it causes the vortex to bow out in the $x$ direction, where the displaced vortex intersects the $x$ axis at $x_d$.   The bent vortex also experiences a Lorentz force due to the component $J_z$, causing the vortex to bend into the shape of a helical arc.  For small values of $J$, the helical distortion is stable, because the inward restoring force per unit length due to the vortex's line tension $\epsilon_\ell$ is able to balance the outward Lorentz force per unit length.  However, when $J$ reaches a critical value, $J_c$, the helical arc becomes unstable and the vortex expands to ever larger radii, allowing it  not only to escape the pinning centers and become depinned but also to grow outward, where it can meet other expanding vortices and undergo flux cutting.  The critical current density $J_c$ can be calculated as follows, using an extension of an approach used in Ref.\ \onlinecite{Pardo07}.

Let us describe the helical arc using the coordinates $(x',y',z')$ with unit vectors $(\hat x',\hat y',\hat z')$, where the $x'$ axis lies along the $x$ axis and the $y'$ and $z'$ axes are tilted relative to the $y$ and $z$ axes by an angle $\phi$ such that the $z'$ axis is parallel to $\bm J$ (see Fig.\ \ref{helixfig}).  Points on the helical arc are described by vector $\bm r' = \hat x' a \cos kz' -\hat y' a \sin kz' + \hat z' z' $.  The origin of the primed frame is chosen to be at $\bm r_0 = (x,y,z) = (x_0,0,0)$, where $x_0 = x_d -a$, such that the coordinates of any point in the primed frame are related to those in the unprimed frame via $(x',y',z') = (x-x_0,y \cos \phi -z \sin \phi, y \sin \phi + z\cos \phi).$ 
In particular, the strong pins at $\bm r_\pm = (0,0,\pm c)$ have the coordinates $\bm r'_\pm = (x'_\pm,y'_\pm,z'_\pm)$  in the primed frame, and since these points lie at the ends of the helical arc, we have $x'_\pm = a-x_d=a \cos (kc \cos \phi)$, $y'_\pm= \mp c \sin \phi = \mp a \sin (kc \cos \phi)$, and $z'_\pm = \pm c \cos \phi$.   

The unit vector tangent to the helical arc in the primed frame is 
\begin{equation}
\hat T'(z') = \frac{\bm r'(z')}{ds'} = \frac{-\hat x' ka \sin kz' -\hat y' ka \cos kz' +\hat z'}{\sqrt{1+k^2a^2}},
\end{equation}
and its derivative is 
\begin{equation}
\frac{d \hat T'(z')}{ds'} =\frac{\hat N'(z')}{\rho_c}=
\frac{k^2 a(-\hat x' \cos kz' +\hat y' \sin kz')}{1+k^2a^2},
\end{equation}
where $\hat N'(z')=-\hat x' \cos kz' +\hat y' \sin kz'$ is the inward-directed unit normal to the helical arc and 
\begin{equation}
\rho_c = \frac{1+k^2a^2}{k^2 a}
\end{equation}
 is the radius of curvature.   

The equation that determines the equilibrium radius $a$ for a given current density is the force-balance equation $\bm f_d + \bm f_r = 0$, where 
the inward restoring force is 
\begin{equation}
\bm f_r = \epsilon_\ell \frac{d \hat T'(z')}{ds'} =\frac{\epsilon_\ell k^2 a\hat N'}{1+k^2a^2},
\end{equation}
and
the outward driving force per unit length is, since $\bm J = J \hat z'$, 
\begin{equation}
\bm f_d = \bm J \times \phi_0 \hat T'=-\frac{J\phi_0 ka}{\sqrt{1+k^2a^2}}\hat N'.
\end{equation} 
The ratio of the magnitudes of the restoring and driving forces is 
\begin{equation}
R = \frac{\epsilon_\ell k }{J\phi_0 \sqrt{1+k^2a^2}},
\label{R}
\end{equation}
and the two forces are exactly balanced when this ratio is equal to 1.  

In the absence of any pinning centers, Eq.\ (\ref{R}) when $a = 0$ tells us that a long, straight isolated vortex subjected to a  parallel current density $J$ is unstable to the growth of a helical perturbation of any wavevector $k < J\phi_0/\epsilon_\ell$, because then $R < 1$.    For wavevectors $k > J\phi_0/\epsilon_\ell$, on the other hand, it is possible for the restoring and driving forces to be balanced when $a = a_u= \sqrt{(\epsilon_\ell /J \phi_0)^2-1/k^2}$, but this is a point of unstable equilibrium.  For $a > a_u$, $R < 1$,  the outward driving force exceeds the inward restoring force, and the helix grows, while for $a < a_u$, $R > 1$,  the inward restoring force exceeds the outward driving force, and the helix shrinks to zero radius.  Thus a straight vortex is stable against infinitesimal perturbations with wavevector $k > J\phi_0/\epsilon_\ell$.  However,  since fluctuations of all wavevectors $k$ are possible, we see that a long, straight isolated vortex is unstable to the growth of helical perturbations; in other words, in the absence of pinning, both $J_{c\parallel} = 0$ and $J_{c\perp} = 0$.  

Returning to the model of the helical arc stretching between two strong pinning centers, we note that the pins place additional constraints on $k$ and $a$ via the equations $\cos (kc \cos \phi) = 1-x_d/a$ and $\sin (kc \cos \phi)=(c/a) \sin \phi$.
Accordingly, we can write the force ratio as $R(j,\phi,\tilde x_d)=f/j$, where $\tilde x_d = x_d/c$ and 
\begin{eqnarray}
j &=& J/J_{c\perp},\\
J_{c\perp} & = & \epsilon_\ell/\phi_0 c,\\
f(\phi,\tilde x_d) &=& \gamma ka/\sqrt{1+(ka)^2},\label{f}\\
ka &=& \cos^{-1}\Big(\frac{\sin^2\phi - \tilde x_d^2}{\sin^2\phi + \tilde x_d^2}\Big)/\gamma \cos \phi,\\
\gamma & = &c/a = 2\tilde x_d/(\sin^2\phi + \tilde x_d^2).
\end{eqnarray}
In the limits $\phi \to 0$ and $\phi \to \infty$,
\begin{eqnarray}
f(0,\tilde x_d) &=& \pi/\sqrt{1+(\pi \tilde x_d/2)^2},\\
f(\pi/2,\tilde x_d) &=&  2\tilde x_d/(1 + \tilde x_d^2).
\end{eqnarray}

\begin{figure}
\includegraphics[width=8cm]{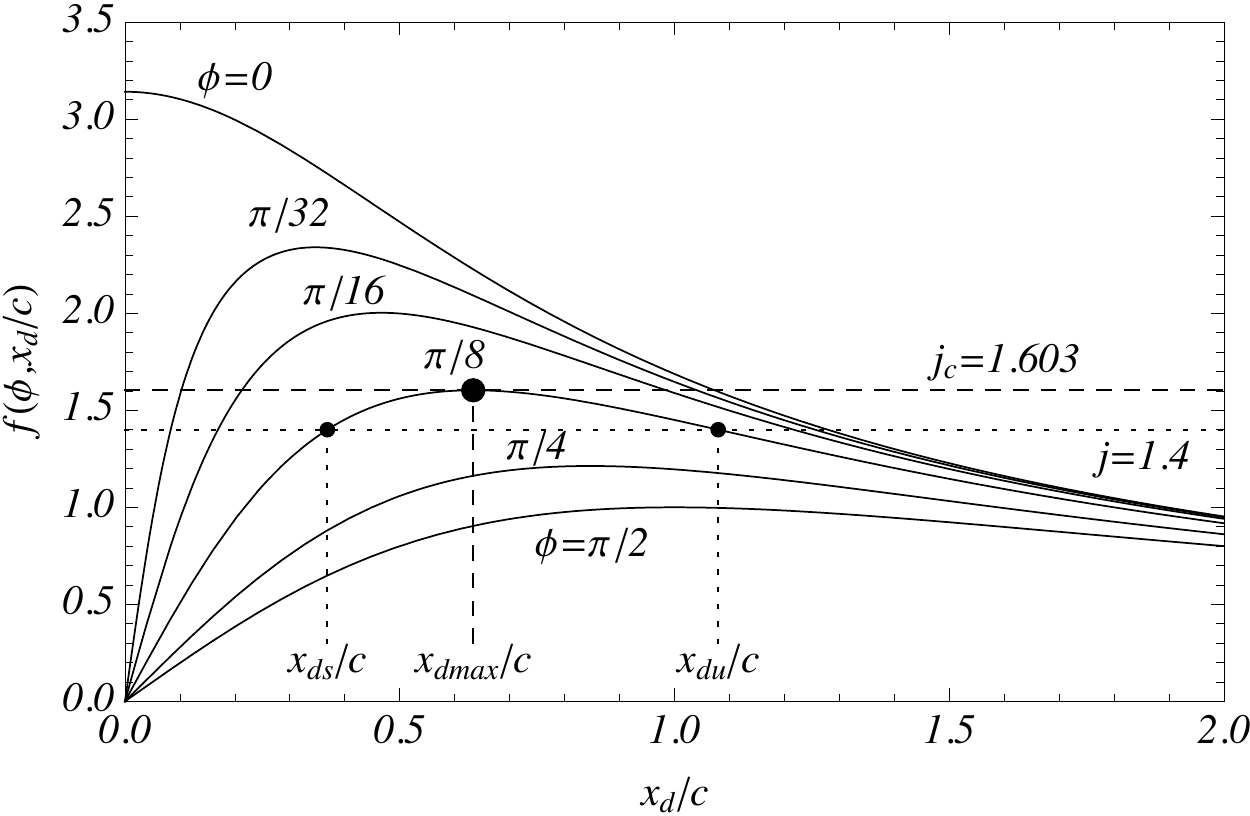}
\caption{ $f$ [solid, Eq.\ (\ref{f})] vs $\tilde x_d = x_d/c$ for $\phi = 0,$ $ \pi/32,$ $ \pi/16,$ $\pi/8,$ $\pi/4,$ and $\pi/2$.  For $\phi = \pi/8$, the forces are balanced $(R = 1)$ for $j = 1.4$ (dashed) at two values of $x_d/c$ (black points), a stable equilibrium point at $x_{ds}/c= 0.368$ and an unstable equilibrium point at $x_{du}/c= 1.080$.  As $j$ increases, the two equilibrium points merge at the maximum of $f$, where $j = j_c$ and $J = J_c$.  For $\phi = \pi/8$, this occurs at $x_{dmax}/c= 0.634$ (large black point), where the dimensionless critical current density is $j_c(\phi) = 1.603$.}
\label{fvsxdplot}
\end{figure}

As illustrated in Fig.\ \ref{fvsxdplot}, when $j$ is not too large, the restoring and driving forces are balanced ($R = 1$) at two displacements $x_d$ of the vortex arc along the $x$ axis, a stable point $x_{ds}$ and an unstable point $x_{du}$.  Note that for $x_d$ slightly larger than $x_{ds}$, $R > 1$ and the inward restoring force exceeds the outward driving force, such that $x_d$ is driven back down to $x_{ds}$, but for $x_d$ slightly smaller than $x_{ds}$, the opposite is true, $R < 1$ and $x_d$ is driven back up to $x_{ds}$.  On the other hand, for $x_d$ slightly larger than $x_{du}$, $R < 1$ and  the outward driving force exceeds the inward restoring force, such that $x_d$ is driven to ever larger values, while for $x_d$ slightly smaller than $x_{du}$, the opposite is true, $R > 1$ and $x_d$ is driven down to $x_{ds}$.  As $j$ increases to the value $j_c$, the equilibrium points merge to the point $x_{dmax}$, where $f$ has its maximum value, $f_{max}(\phi) = j_c$ and $R = 1$.  However, for any values of $j > j_c(\phi)$, the force ratio $R = f/j$ is less than 1, indicating that the outward driving force exceeds the inward restoring force for all values of $x_d$, and the helical arc must expand to ever-increasing displacements.  Thus when $j > j_c(\phi)$ or $J > J_c(\phi)$, the vortex arc undergoes a helical instability that not only leads to flux-line cutting with other vortices in the sample but also  allows the free portion of the vortex to escape from the strong pins and undergo flux flow.

\begin{figure}
\includegraphics[width=8cm]{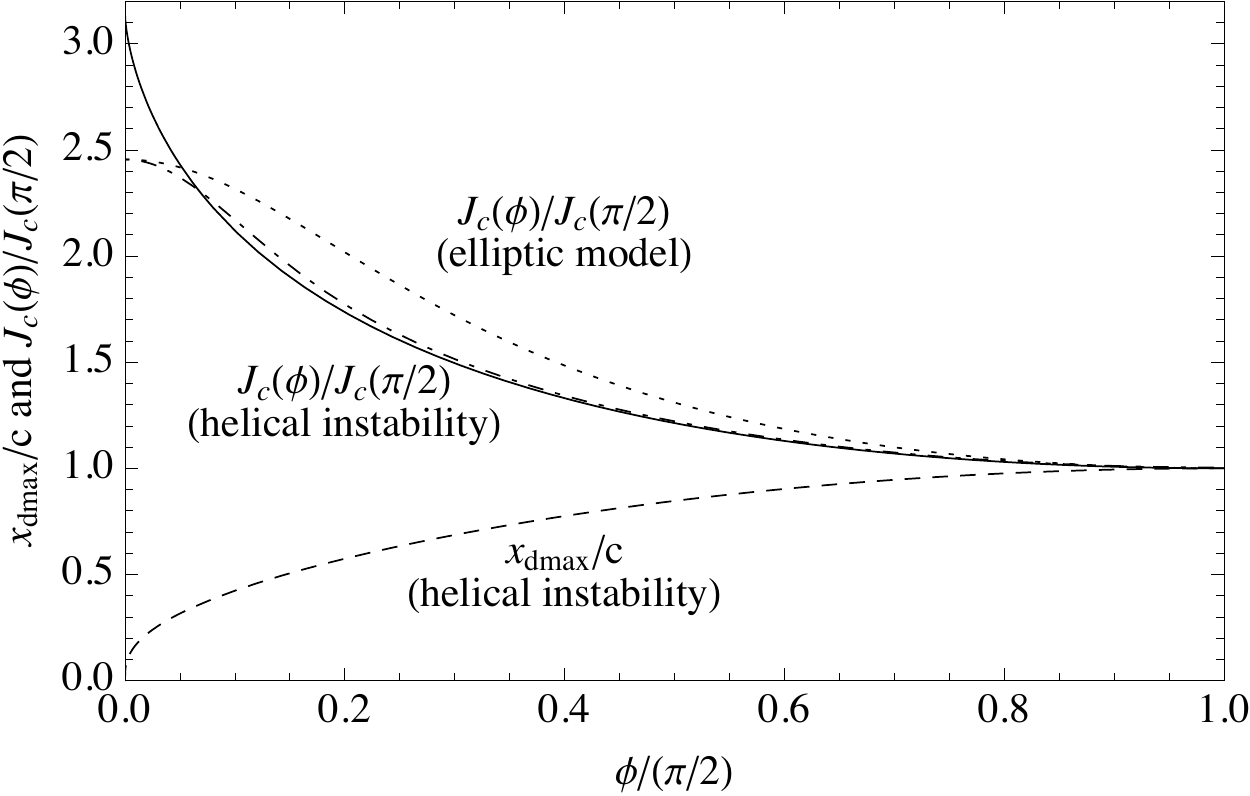}
\caption{ $J_c(\phi)/J_c(\pi/2)$ (solid) and $x_{dmax}/c$ (dashed) vs $\phi/(\pi/2)$ obtained using the helical-instability calculation described in the text, for which $J_c(\pi/2) = J_{c\perp} = \epsilon_\ell/\phi_0 c$ and $J_c(0) = J_{c\parallel} =  \pi\epsilon_\ell/\phi_0 c$.  The dot-dashed curve shows the pin-distribution average (\ref{Jcavg}) obtained by convolving the helical-instability $J_c(\phi)$ with the gaussian of Eq.\ (\ref{gaussian}) with $\Delta \phi = \pi/20,$ which yields $J_c(0)/J_c(\pi/2) = 2.455.$ For comparison, the dotted curve shows the angular dependence of the elliptic model [Eq.\ (\ref{Jcellipse})] for the same $J_{c\parallel}/J_{c\perp} = 2.455$.  }
\label{Jcxdmaxplot}
\end{figure}

Results for $J_c$ and $x_{dmax}$ vs $\phi$ obtained using this simple model are shown in Fig.\ \ref{Jcxdmaxplot}.  The key features are that (a) $J_{c\parallel}$ is proportional to $J_{c\perp}$, consistent with the experimentally observed correlation between these critical currents, (b) the same helical instability leads to both depinning and flux cutting, showing that there is an intimate connection between these processes, and (c)  the threshold
$J_{cd}$ for depinning is reduced to zero as the magnitude of 
$J_{\parallel}$ increases to its maximum threshold value
$J_{c\parallel}$ while the threshold
$J_{cc}$ for flux-line cutting is reduced to zero as 
$J_{\perp}$ increases to its maximum threshold value 
$J_{c\perp}$. 

On the other hand, deficiencies of this model are that (a) the ratio $J_{c\parallel}/J_{c\perp} = \pi,$ but experimentally this ratio is apparently sample- and field-dependent, (b) it is an isolated-vortex model, which ignores the intervortex interactions that are generally important in type-II superconductors, (c) it does not reproduce the experimentally observed angular dependence of $J_c(\phi)$, which is often well described using the elliptic model, and (d) the pinning model used here is greatly oversimplified and would need  extensions to account for more realistic distributions of pinning centers.  The latter effect can be crudely estimated by convolving the helical-instability $J_c(\phi)$ obtained above with a gaussian distribution function to obtain a pin-distribution average,
\begin{eqnarray}
J_{c,avg}(\phi) &=& \int_{-\pi}^\pi g(\phi-\psi)J_c(\psi) d\psi,
\label{Jcavg} \\
g(\psi)&=&(1/\sqrt{\pi}\Delta \phi)\exp(-\psi^2/\Delta \phi^2),
\label{gaussian}
\end{eqnarray}
where $\Delta \phi$ is a measure of the width of the distribution of the pin-to-pin vectors around the $z$ axis.
$J_{c,avg}(\phi)$ is plotted for $\Delta \phi = \pi/20$ as the dot-dashed curve in Fig.\ \ref{Jcxdmaxplot}, where a corresponding plot of the elliptic model is shown as a dotted curve.  

\section{Discussion\label{Discussion}}

\subsection{Relation to experiments}

The theoretical calculations of Sec.\ \ref{calcs} relate most closely to pioneering experiments by Walmsley.\cite{Walmsley72a}  His experiments were carried out using a cylinder of Pb-40 at\% Tl alloy, subjected to combinations of a magnetic field $H_a$ and electrical current $I$, while both the axial magnetization and the axial resistance were simultaneously monitored.  As $H_a$ was held fixed, a paramagnetic axial magnetization was found to develop as the current $I$ increased above the critical current $I_c$ and flux flow occurred.  This corresponds qualitatively to the behavior  when $r < 1$ discussed in Sec.\ \ref{para} and shown in Fig.\ \ref{HJPlotreq0p5}.  The experimentally observed behavior can be interpreted as resulting from the nucleation of helical vortices produced by the combination of the applied longitudinal field and the self-field.  The helical vortices move toward the center of the cylinder, building up a longitudinal paramagnetic moment.  As discussed in Sec.\ \ref{Bconsumption}, there is a balance between flux transport and flux cutting, but when $r < 1$, which corresponds to a low efficiency of flux cutting for reduction of $B$, a paramagnetic longitudinal moment is produced.  

The experiments\cite{Walmsley72a} also revealed that when the applied magnetic field was reduced to a nominally zero value, a spontaneous longitudinal magnetic moment of either sign developed when $I > I_c$.  Our interpretation of this effect is that, although the longitudinal field was adjusted to a very small value, it could not be made exactly zero, and as a result helical vortices penetrated from the surface during flux flow, carrying longitudinal magnetic flux toward the center.  Figure \ref{Hzvsrhoparamagplot} shows calculations for $r < 1$ revealing large positive values of $H_z$ in the interior of the sample even for very small positive values of $H_z = H_a$ at the surface.  Similarly, large  negative values of $H_z$ in the interior of the sample would also be expected theoretically for very small negative values of $H_z = H_a$ at the surface.

The theoretical calculations of Sec.\ \ref{Bdependence} provide a qualitative explanation of experiments by Karasik and Vereshchagin\cite{Karasik71}  showing striking differences between the critical currents of Ti-22 at\% Nb and Ti-36 at\% Nb wires measured in longitudinal and transverse magnetic fields. The critical currents in longitudinal fields $H_a$ were all observed to increase initially with $H_a$ and to exhibit high peaks, where the longitudinal critical current was one or two orders of magnitude larger than the transverse critical current at the same field.  Such behavior can be understood qualitatively as discussed in Sec.\ \ref{Bdependence} and shown in Fig.\  \ref{IcvsHaKimplot}.

Various researchers experimentally studying the electric field in longitudinal geometry have found inhomogeneities in $\bm E $ along the sample length.\cite{Walmsley72a,Irie74,Irie75,Ezaki76,Cave78,Matsushita98}  However, such effects cannot be understood in terms of the present theory, which assumes uniformity of $\bm E$ just above the critical current.

\subsection{Force-free configurations}

In analyzing his experiments  revealing a paramagnetic moment in current-carrying on Pb-40 at\% Tl alloy cylinders subjected to a parallel magnetic field, Walmsley\cite{Walmsley72a} compared his results with force-free field theory, in which $\bm J$ is parallel to $\bm H$.  Using the terminology of the present paper, in cylindrical geometry force-free fields are those for which $J_\perp = 0$.  After setting $J_\perp = 0$, we can integrate  Eq.\  (\ref{Jperp}) to obtain\cite{Clem76}
\begin{eqnarray}
H(\rho) &=& H(0)\exp[-\int_0^\rho r^{-1}\sin^2\alpha(r) dr],\;{\rm or} \\
 &=& H(R)\exp[-\int_\rho^R r^{-1}\sin^2\alpha(r) dr].
\end{eqnarray}
Since these equations hold for any arbitrary function $\alpha(\rho)$, there is an infinite number of possible force-free solutions.  For each $\alpha(\rho)$, the corresponding $J_\parallel(\rho)$ can be calculated from Eq.\ (\ref{Jparallel}).

Walmsley\cite{Walmsley72a} focused his attention on Bessel-function solutions, which follow from the assumption that $J_\parallel(\rho) = k H(\rho)$, where [see Eq.\ (\ref{Jparallel})]
\begin{equation}
k = \frac{d\alpha}{d\rho} + \frac{\sin\alpha \cos\alpha}{\rho}
\end{equation}
is a constant.  This equation is satisfied when\cite{Clem76} 
\begin{equation}
\alpha(\rho) = \tan^{-1}[J_1(k\rho)/J_0(k\rho)],  
\end{equation}
such that 
\begin{eqnarray}
H_z(\rho)& =& H(0)J_0(k\rho), \\
H_\theta(\rho)& =& H(0)J_1(k\rho), \\
H(\rho)& =& H(0)\sqrt{J_0^2(k\rho)+J_1^2(k\rho)}.
\end{eqnarray}
Since $0 < H_z(R) < H(0)$ and $I = 2\pi R H_\theta(R) > 0$ when $0< kR < 2.4048$, solutions for $k$ in this range yield paramagnetic force-free solutions for a cylinder of radius $R$ carrying a current $I$ in the $z$ direction in a parallel field $H_a = H_z(R).$  These solutions were found to be in good agreement with many of the  measurements.\cite{Walmsley72a}

However, the above Bessel-function solutions are only one of infinitely many possible force-free solutions, and all such solutions fail to describe the full physics of the dynamical processes occurring at the critical current $I_c$ of a superconducting cylinder in a parallel applied field.  Force-free solutions do not satisfactorily explain the origin of the longitudinal electric field that appears above $I_c$. If we had a truly force-free situation where $J_\perp = 0$ and $J_\parallel > 0$, then we would also have $E_\perp = 0$ and $E_\parallel > 0$, which violates Faraday's law.  As a consequence, it would be impossible to balance the transport and cutting contributions to $\partial B/\partial t$ appearing in Eq.\  (\ref{dBdt}).

\subsection{Interactions between flux cutting and flux depinning}

Further theoretical work should be done to examine flux cutting in the presence of pinning centers.  What initiates flux cutting evidently is the helical expansion instability first discussed in a geometries with linear dimensions of the order of the penetration depth $\lambda$,\cite{Clem77,Genenko94a,Genenko94b,Genenko95a,Genenko95b,Genenko96} such that interactions with the surface could stabilize the vortex against the helical expansion instability for small currents.  However, Brandt\cite{Brandt81a,Brandt81b} showed that, because of their collective behavior,
an array of vortices parallel to the axis of a cylinder of radius much larger than the intervortex spacing is unstable to a collective helical expansion instability in arbitrarily small longitudinal currents.  These results suggest that pinning centers help to stabilize the vortex array against the helical expansion instability.  This is consistent with the experimental observation that the longitudinal and transverse critical currents are roughly proportional to each other.

The intimate relationship between flux-line cutting and flux depinning is also worthy of deeper study.  Numerous experiments have revealed that the critical-current densities in longitudinal and transverse applied fields usually are roughly of the same order of magnitude, although the critical current in a longitudinal field is generally somewhat higher than that in a transverse field.  This suggests that the thresholds for flux-line cutting and flux depinning are closely linked.  The model described in Sec.\ \ref{HelicalArcs}, in which the same helical instability simultaneously initiates flux cutting and flux transport, should point the way to deeper understanding of this interrelationship.

\subsection{Accounting for both flux cutting and flux depinning in time-dependent problems}

While the present paper has dealt only with a steady-state problem in which there is no time dependence of the time-averaged quantities $\bm B$, $\bm H$, and $\bm E$, it should be possible to extend the above approach to solve  problems in which these quantities are time-dependent at frequencies of interest to power applications.  The equations needed are simplest when the superconductor is macroscopically isotropic, i.e., when the penetration depth $\lambda$ and coherence length $\xi$ are the same along different crystallographic directions, and the critical current densities for flux depinning and flux cutting depend only upon the magnitude of $\bm B$ and the angle $\phi$ between $\bm J$ and $\bm B$.  The basic equations are then
\begin{eqnarray}
&&\bm J = \nabla \times \bm H, \label{Ampere}\\
&&\nabla \times \bm E = -\partial \bm B/\partial t, \label{Faraday}
\end{eqnarray}
where $\bm H = \hat H H$, $H(B) = (1/\mu_0)\nabla_B F(B)$, and $F(B)$ is the Helmholtz free energy density in the superconducting state.\cite{Fetter69}  The displacement current can safely be neglected at low frequencies.   $H(B)$ can be obtained using standard magnetization measurements. Introducing the unit vector $\hat \alpha = \bm B/B = \bm H/H$, we can define the component of $\bm J$ along $\bm B$ as $\bm J_\parallel = \hat \alpha J_\parallel$.  The component of $\bm J$ perpendicular to $\bm B$ is then $\bm J_\perp = \bm J -\bm J_\parallel$.  Similarly, the components of $\bm E$ parallel and perpendicular to $\bm B$ are $\bm E_\parallel = \hat \alpha E_\parallel$  and $\bm E_\perp = \bm E -\bm E_\parallel$.  What is also needed, but seldom experimentally determined to date, are the dependencies of $\rho_\parallel$ and $\rho_\perp$ in the expressions
\begin{eqnarray}
\bm E_\parallel&=&\rho_\parallel \bm J_\parallel, \\
\bm E_\perp&=& \rho_\perp \bm J_\perp,
\end{eqnarray}
bearing mind that in type-II superconductors each of the quantities $\rho_\parallel$ and $\rho_\perp$ depends strongly upon its corresponding current density, with a well-defined increase when its threshold value ($J_{cc} = J_c|\cos \phi|$ or $J_{cd}=J_c|\sin \phi|$, as in Fig.\ \ref{EllipticPlot}) is exceeded.  Finally, what is needed is a model for the behavior of $J_c(B,\phi)$, similar to the elliptic critical-state model of Eq.\  (\ref{Jcellipse}), as shown in Fig.\ \ref{EllipticPlot}, where $\phi$ is the angle between $\bm J$ and $\bm B$.  In many practical cases, it is likely that the time evolution of the magnetic-field and current-density profiles will need to be determined numerically by solving Eq.\  (\ref{Faraday}) step by step in time, as has been done in simpler cases in Refs.\ \onlinecite{Perez85a} and \onlinecite{Romero03a}.

Further extensions of the above procedure would be necessary to incorporate the effects of anisotropy in both the intrinsic properties [i.e., if  $\lambda$, $\xi$, and $\bm H(\bm B)$ differ along different crystallographic directions] and extrinsic properties (e.g, if the pinning centers have linear character and are aligned along different crystallographic directions).  

\begin{acknowledgments}
I thank A. M. Campbell, J. H. Durrell, V. G. Kogan, W. K. Kwok, A. Malozemoff, M. Weigand, and U. Welp for stimulating questions and suggestions.
 This research, supported in part by the U.S. Department of
Energy, Office of Basic Energy Science, Division of Materials
Sciences and Engineering, was performed at
the Ames Laboratory, which is operated for the U.S. Department
of Energy by Iowa State University under Contract No.
DE-AC02-07CH11358.  This research also was
supported in part  by the Center for Emergent Superconductivity, an Energy Frontier Research Center funded by the U.S. Department of Energy, Office of Science, Office of Basic Energy Sciences under Award Number DE-AC0298CH1088.

\end{acknowledgments}

\end{document}